\documentclass[twoside,slac_one]{revtex4}
\usepackage{graphicx}
\usepackage{fancyhdr}
\usepackage{amsmath} 
\usepackage{bm}
\usepackage{amsxtra}
\usepackage{amssymb}
\usepackage{amsthm}
\usepackage{latexsym}
\usepackage{lscape}

\newcommand{\pT}{$p_{T}$}
\newcommand {\pp} {{\it pp}}
\begin{document}

\title{Jets and Jet-like Correlations at RHIC}

%

\author{Helen Caines}
\affiliation{Department of Physics, Yale University, New Haven, CT, USA}

\begin{abstract}
I present an overview of some of the recent results on jets and jet-like correlation measurements from the Relativistic Heavy-Ion Collider (RHIC) at Brookhaven National Laboratory. 

Jets are produced in the initial hard scatterings of an event and can therefore be exploited as probes of the hot and dense medium produced in heavy-ion collisions. Previous RHIC results indicate that this medium, the Quark Gluon Plasma (sQGP), is strongly coupled, with partonic degrees of freedom. High \pT\  colored partons passing through the sQGP are therefore believed to suffer energy loss via induced gluon radiation and elastic collisions, before exiting the medium and fragmenting in vacuum. Jet reconstruction and high \pT\ correlation studies allow us to investigate how the partons interact with the medium and how the medium responds to the partons moving through it. By comparing measurements from \pp\ and {\it d}-Au to those in Au-Au collisions at $\sqrt{s_{NN}}$ = 200 GeV we aim to disentangle cold nuclear matter effects from those of the hot and dense sQGP.

\end{abstract}

\maketitle

\thispagestyle{fancy}


\section{Introduction}

Over the past decade the RHIC experiments have produced significant evidence that a Quark Gluon Plasma (sQGP) is being produced  in  ultra-relativistic heavy-ion collisions. This sQGP is strongly coupled and has partonic degrees of freedom. Hard  probes are now being used to study  how partons interact with the medium, and how the sQGP responds to energy deposited by these highly energetic partons as they travel  through it. The results from Au-Au collisions are compared to those from \pp\ and  {\it d}-Au collisions where no QGP is believed to be created.

Jet quenching, the loss of energy of hard scattered partons to the medium, was first observed at RHIC via  the single particles nuclear modification factor, $R_{AA}$~\cite{Raa}. The nuclear modification factor is defined as the ratio of Au-Au \pT\ spectrum normalized to the number of binary collisions ($N_{bin}$) to that  of  the \pp\ data measured at the same collision energy.  The $R_{AA}$ of high \pT\ particles in central Au-Au indicated a  suppression of up to a factor of 5~\cite{Raa}, while  photons, colorless objects, reveal an  $R_{AA}$ =1~\cite{PhenixPhoton}. These results suggest that the suppression is due to partons interacting with, and hence loosing energy to,  the hot and dense colored medium.   $R_{AA}$ measures however have a couple of  limitations. First it is not possible to infer the initial partonic energy from the final state hadron. Second surface emission leads to an inherent insensitivity to the medium's density; no matter how dense  the medium those partons near the surface will always escape and be detected~\cite{SurfaceBias}.

To attempt to alleviate both these issues the collaborations are now performing  full jet reconstruction. An added advantage is that by studying the fragmentation patterns we can hope determine not only how much energy is lost from the initial parton but also how the energy is redistributed. Both STAR and PHENIX use jet finding algorithms from the FastJet package~\cite{FastJet}, in addition PHEINX  uses a Gaussian filter code, details of which can be found here~\cite{GausFilt} and STAR has used a traditional mid-point cone algorithm~\cite{Cone}.

Neither RHIC experiment has hadronic calorimetry included in their detector designs, hence the full jet energy is not directly accessible but is assessed by combining  charged particle momenta via tracking devices with neutral energy measurements from electro-magnetic calorimetry. In addition the energy from long-lived neutral hadrons, such as the neutron and $K^{0}_{L}$ is missed. This leads to a significant difference in the jet energy reconstructed and that of the initial, so-called particle level jet. This difference, as well as the jet energy resolution, has to be evaluated and corrected for before final results can be presented. For instance, the detector performances have been evaluated via simulations by STAR and show that the jet energy resolution in \pp\ data  varies from 10-25$\%$, for 40-10 GeV/{\it c} jets~\cite{CainesQM}.

\section{Jets in pp and d-Au collisions}

\begin{figure}[htb]
	\begin{minipage}{0.46\linewidth}
		\begin{center}
			\includegraphics[width=0.7\linewidth,angle=90]{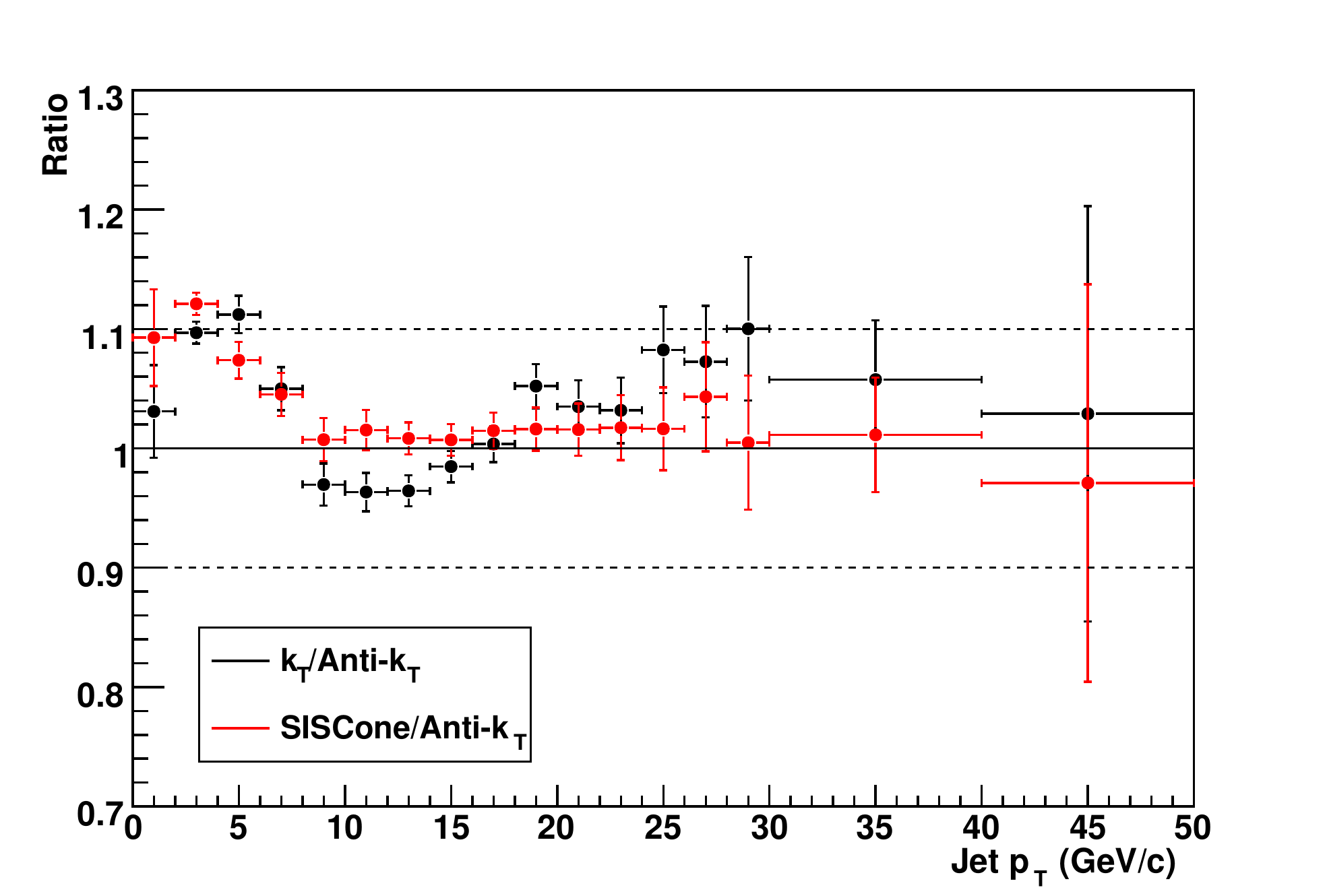}
		\end{center}
	\end{minipage}

	\caption{The ratio of reconstructed jets for various FastJet algorithms, K$_{T}$/Anti-K$_{T}$ and SISCone/Anti-K$_{T}$ as a function of reconstructed jet p$_{T}$.}
			\label{Fig:JetCompare}
\end{figure}

\begin{figure}[htb]
	\begin{minipage}{0.46\linewidth}
		\begin{center}
			\includegraphics[width=0.7\linewidth]{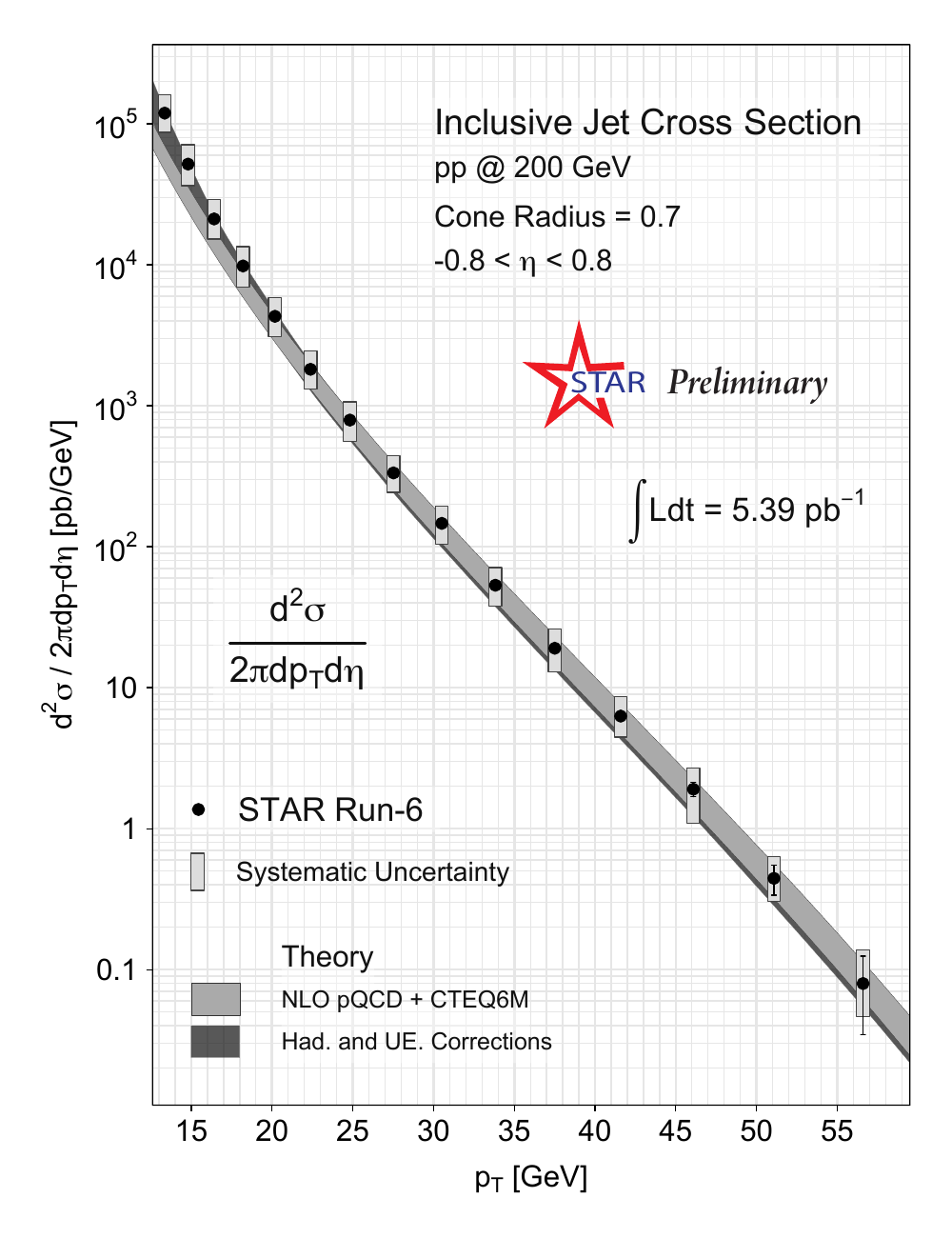}
		\end{center}
	\end{minipage}
	\begin{minipage}{0.46\linewidth}
		\begin{center}
			\includegraphics[width=0.7\linewidth]{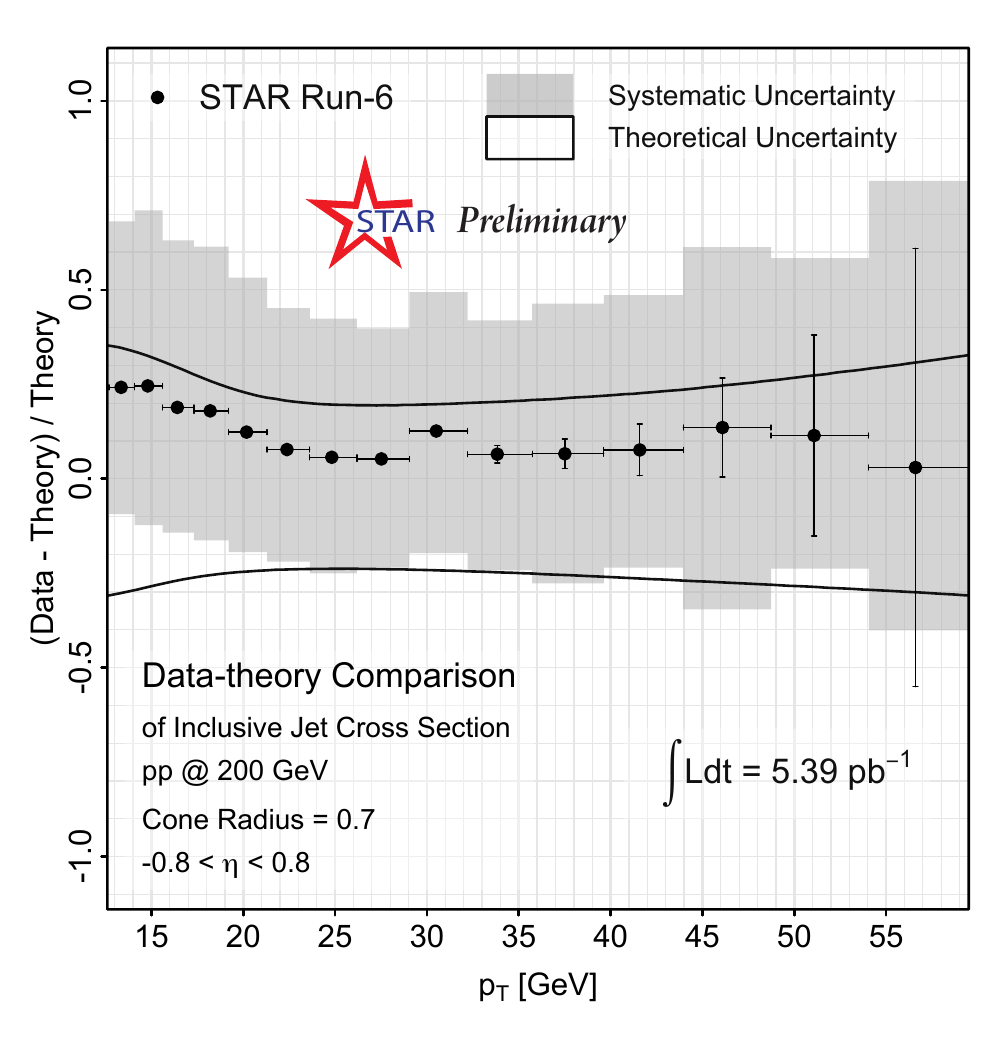}
		\end{center}
	\end{minipage}
	\caption{Left: The 2006 measured inclusive jet cross section for \pp\ collisions at  $\sqrt{s_{NN}}$=200 GeV. Right: (Data-Theory)/Theory.}
			\label{Fig:JetXSec}
\end{figure}

\begin{figure}[htb]
	\begin{minipage}{0.46\linewidth}
		\begin{center}
			\includegraphics[width=0.7\linewidth]{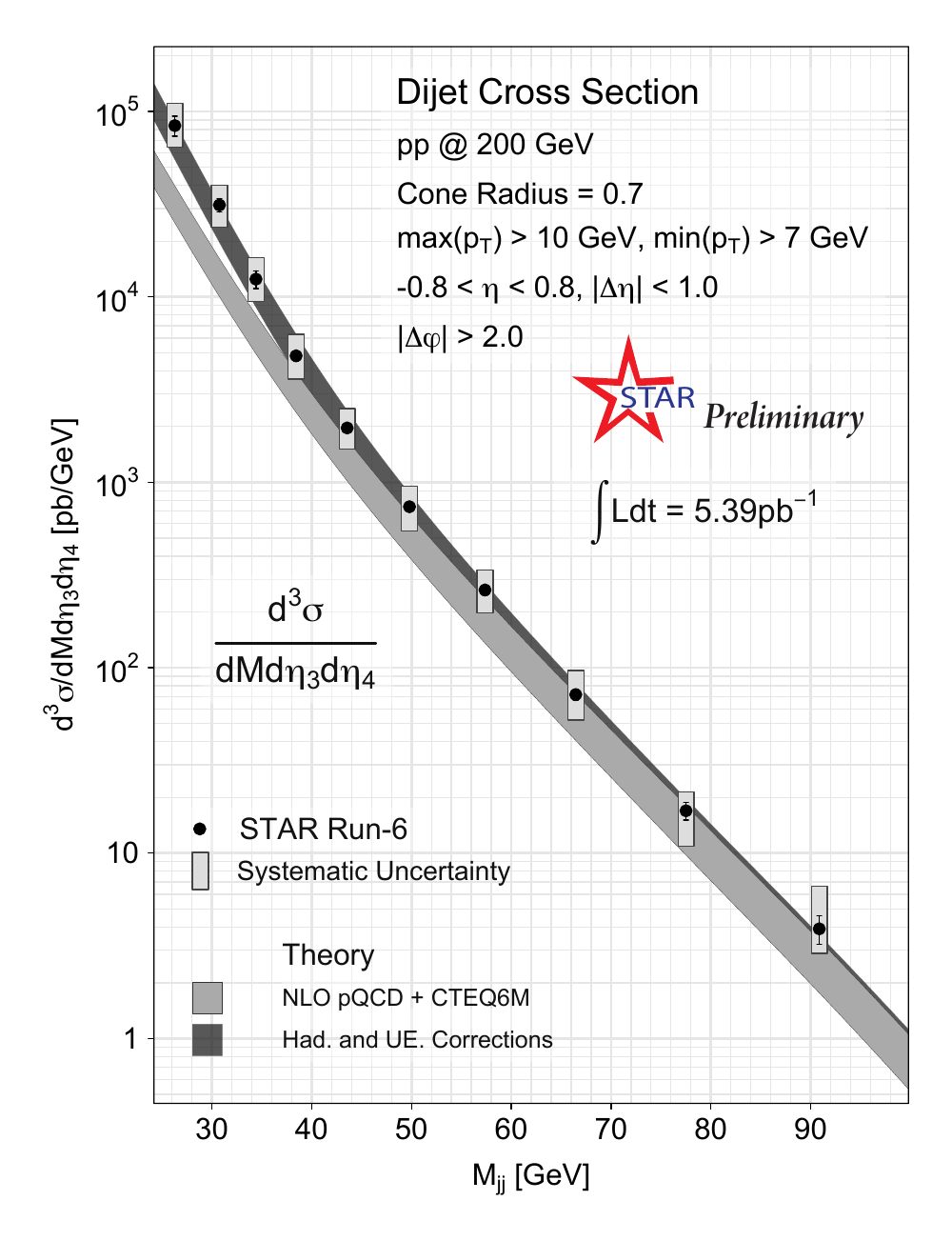}
		\end{center}
	\end{minipage}
	\begin{minipage}{0.46\linewidth}
		\begin{center}
			\includegraphics[width=0.7\linewidth]{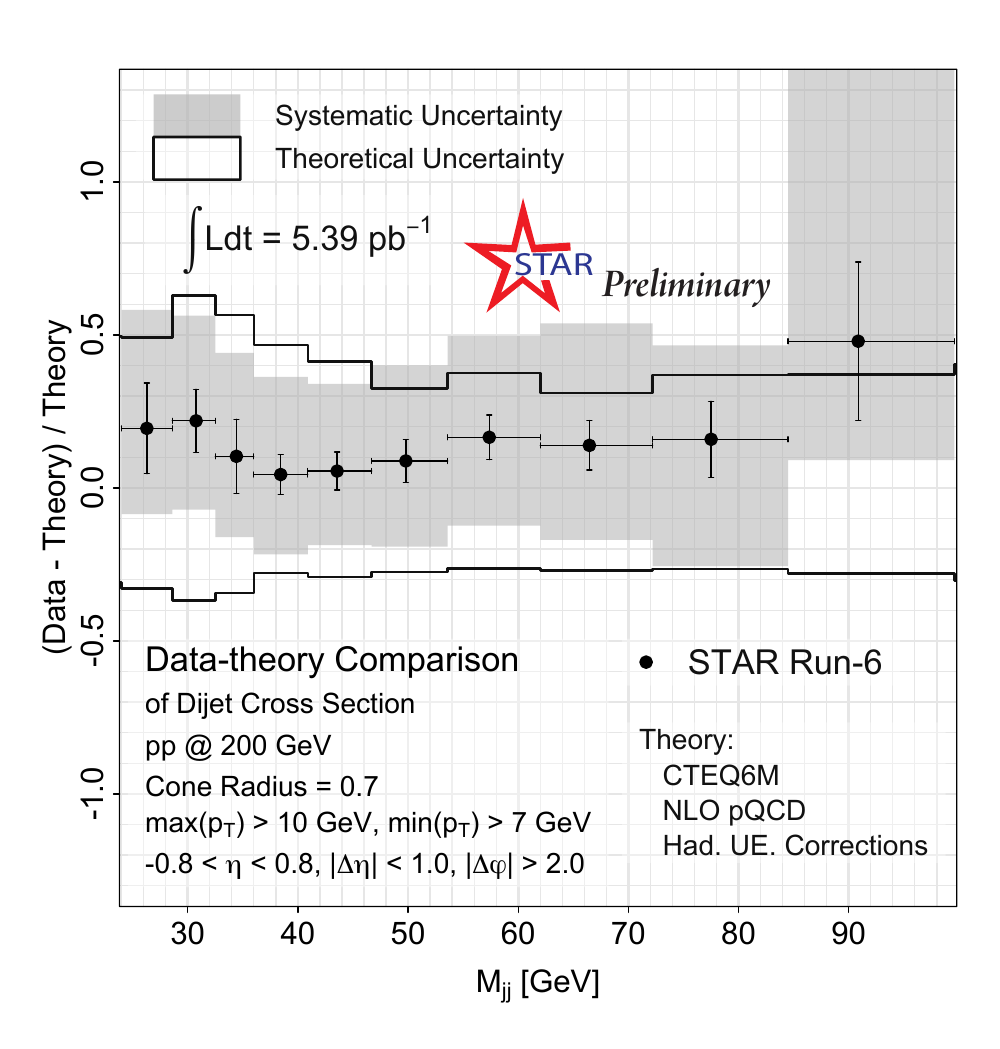}
		\end{center}
	\end{minipage}
	\caption{Left: The 2006 measured inclusive di-jet cross section for \pp\ collisions at  $\sqrt{s_{NN}}$=200 GeV. Right: (Data-Theory)/Theory.}
			\label{Fig:DiJetXSec}
\end{figure}

\noindent  For  \pp\ collisions the reconstructed raw jet spectra reconstructed with the Anti-K$_{T}$, K$_{T}$ and SISCone algorithms  were the same within 10$\%$, confirming that they have similar behaviours in this low multiplicity data, Fig.~ref{Fig:JetCompare}~\cite{CainesQM}. The inclusive jet, Fig.~\ref{Fig:JetXSec}, and di-jet cross-sections, Fig.~\ref{Fig:DiJetXSec},  have been measured by STAR using the increased statistics of the 2006 data~\cite{DijetXSect}. A  midpoint cone algorithm~\cite{Cone} with a cone radius of 0.7, a split-merge fraction 0.5 and a seed energy of 0.5 GeV was used. When hadronization and underlying event uncertainties are included both sets of data are well described by NLO theory~\cite{NLO1,NLO2}. 

\begin{figure}[htb]
	\begin{minipage}{0.46\linewidth}
		\begin{center}
			\includegraphics[width=0.7\linewidth]{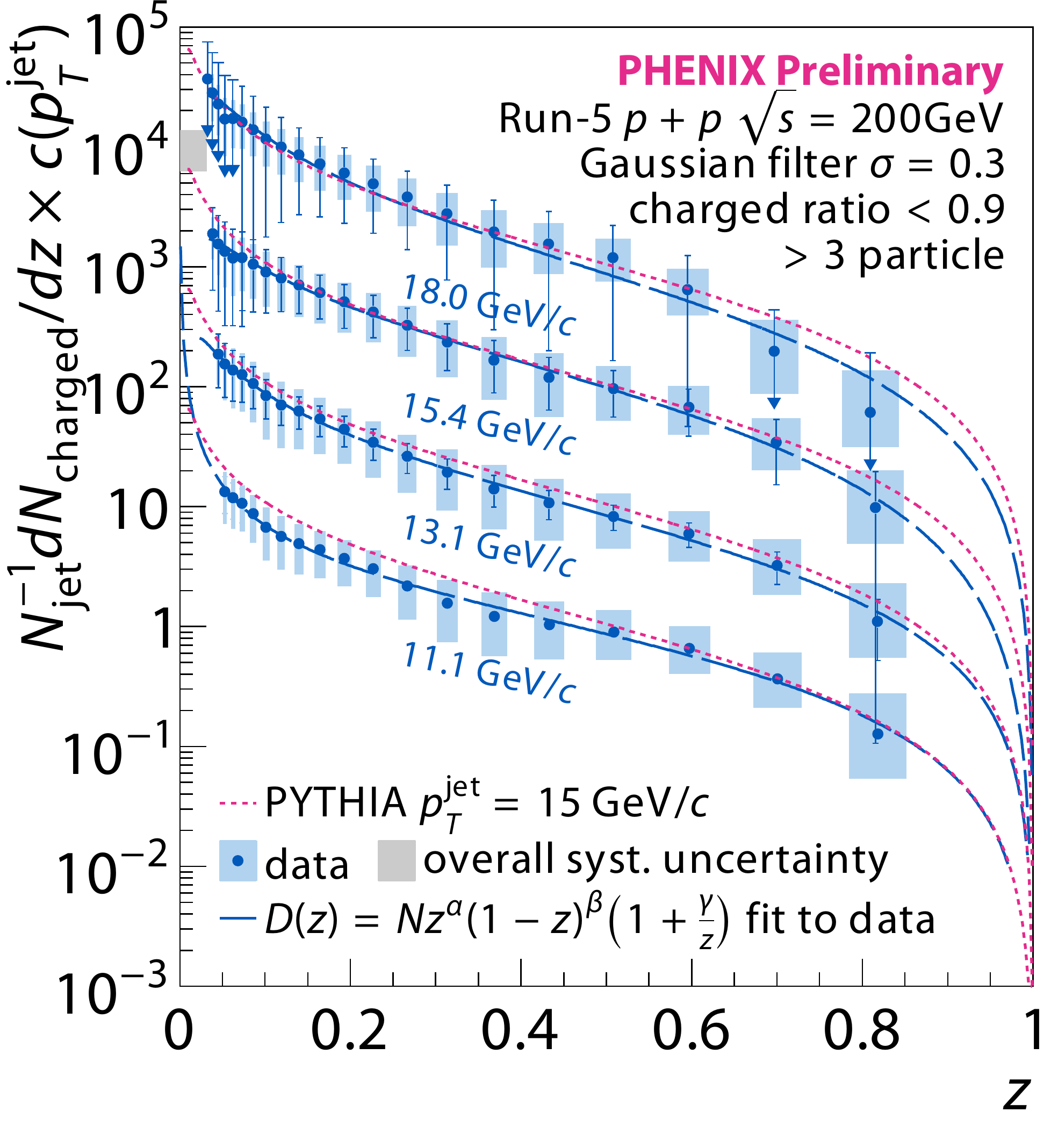}
		\end{center}
	\end{minipage}
	\begin{minipage}{0.46\linewidth}
		\begin{center}
			\includegraphics[width=0.9\linewidth]{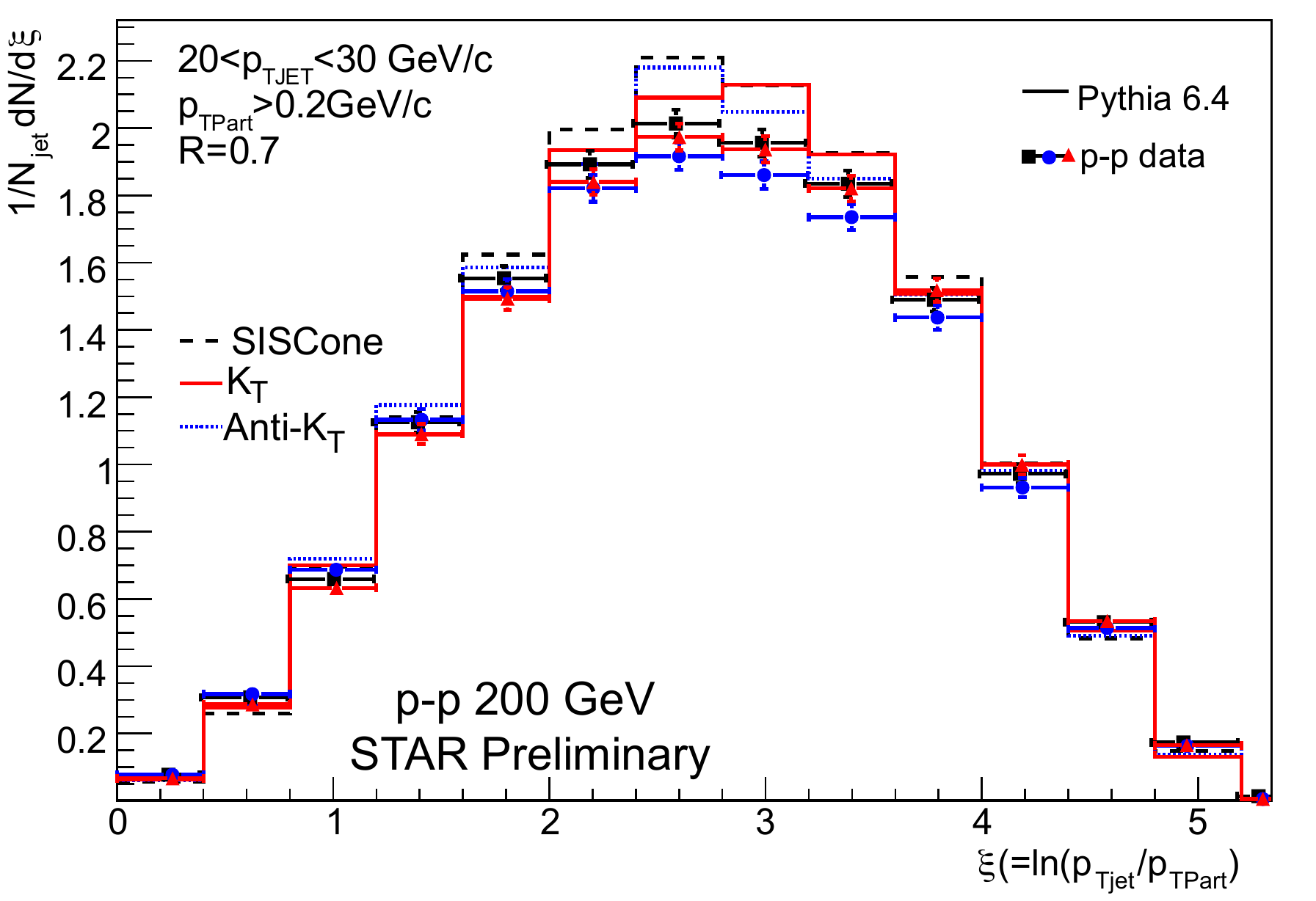}
		\end{center}
	\end{minipage}
	\caption{Left: PHENIX Run-5 \pp\  charged particle (electrons rejected)  fragmentation function as a function a z (=p$_{hadron}$ 
/p$_{jet}$.  A vertical scaling of $c(p_T^\mathrm{jet}) = 10^k, k = 0, \dots,   3$ is applied. The shaded boxes indicate  systematic  uncertainties, and error bars indicate statistical  uncertainties.  From ~\cite{LaiWW}. Right: Jet FF as a function $\xi$ from STAR using the K$_{T}$, Anti-K$_{T}$ and SISCone algorithms for R=0.7. The curves are the PYHTIA predictions. }
			\label{Fig:JetFF}
\end{figure}

The measured fragmentation functions of both STAR~\cite{CainesQM09} and PHEINX~\cite{LaiWW} agree, within errors, with PYTHIA simulations~\cite{PYTHIA}, see Fig.~\ref{Fig:JetFF}. The PHENIX data use the Gaussian filter with a width of 0.3, and  are fully corrected. While the data from STAR are not yet corrected to the particle level and are therefore compared to PYTHIA 6.410~\cite{PYTHIA}, tuned to the CDF 1.96 TeV data (Tune A), predictions passed through STAR's simulations and reconstruction algorithms. The agreement with PYTHIA simulations even for R=0.7 suggests that there are only minor NLO contributions beyond those mimicked in the PYTHIA parton-shower calculations at RHIC energies.

\begin{figure}[htb]
	\begin{minipage}{0.46\linewidth}
		\begin{center}
			\includegraphics[width=1.1\linewidth]{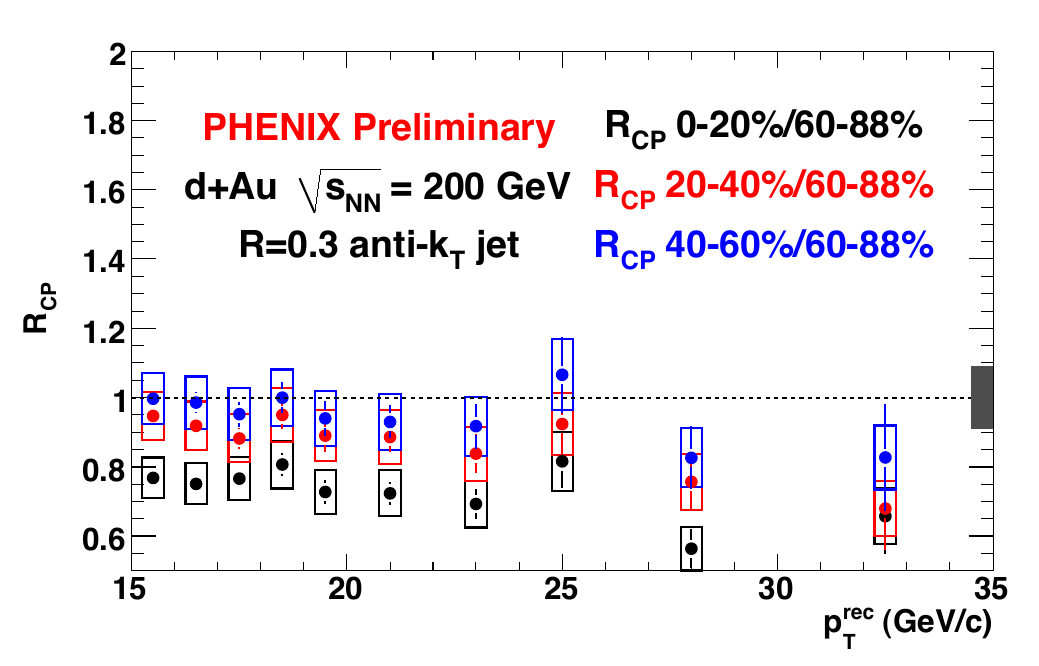}
		\end{center}
	\end{minipage}
	\begin{minipage}{0.46\linewidth}
		\begin{center}
			\includegraphics[width=0.9\linewidth]{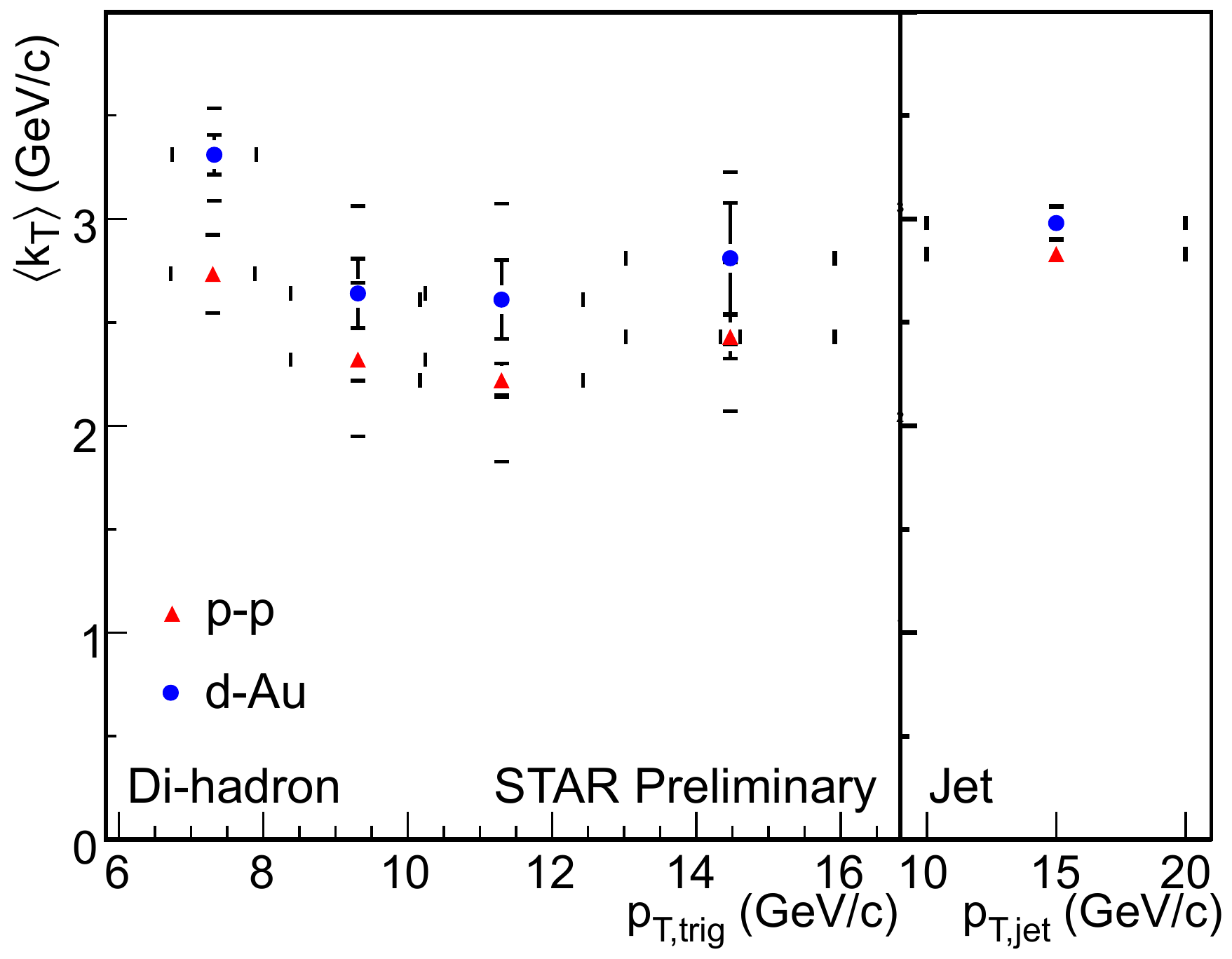}
		\end{center}
	\end{minipage}
	\caption{Left : Jet $R_{CP}$ for d-Au collisions for three different centrality selections. Right: The measured $\langle k_{T} \rangle$  for  {\it d}-Au and  \pp\ data at $\sqrt{s_{NN}}$=200 GeV from di-hadron correlations and di-jet measurements from STAR. Vertical bars show the statistical and systematic uncertainties, horizontal bars indicate the bin widths.}
			\label{Fig:dAuJet}
\end{figure}

We also investigate jet production in  {\it d}-Au collisions, where cold nuclear mater effects are expected to be present but no QGP formed.  For example, the presence of the Au nucleus may  induce  additional initial and final state radiation, or result in scatterings of fragmentation particles as they escape the nucleus. The measured  {\it d}-Au mid-rapidity jet nuclear modification factor, $R_{CP}$ , where peripheral {\it d}-Au events are used instead of \pp\ data, is shown from PHENIX in  Fig.~\ref{Fig:dAuJet} left panel, for three different  centrality bins. Here the Anti-K$_{T}$ algorithm with R=0.3 was used. A slight modification of the jet spectrum is observed in {\it d}-Au collisions, with the central {\it d}-Au jet cross-section showing the greatest suppression. These results are consistent with the $\pi^{0}$ results and  are likely an   indication of cold nuclear matter effects such as modifications of the nuclear PDFs and/or energy loss in the cold matter. Since  these effects may  result  in  more subtle modifications  than that of the overall jet yields, another way to probe for  these cold nuclear matter effects is via di-jet correlations. Re-scatterings in the nucleus my result in a broadening of the di-jet $\Delta \phi$ distribution. The mean transverse momentum of the fragmentation products with respect to the jet axis, $\langle  j_{T} \rangle$, and  the mean transverse momentum kick given to  di-jet pair, $\langle k_{T} \rangle$, are two variables used for these investigations. The  $\langle j_{T} \rangle$ was measured via di-hadron correlations and found to be constant at $\approx$0.55 GeV/{\it c} for all \pT\ triggers  measured and for both \pp\ and  {\it d}-Au events at  $\sqrt{s_{NN}}$=200 GeV~\cite{Mondal}. However, as shown in  Figure~\ref{Fig:dAuJet} right panel  the $\langle k_{T} \rangle$  is systematically higher for the  {\it d}-Au data than for the \pp\ data for all \pT\ jet and \pT\ trigger ranges measured. This suggests that while cold nuclear matter effects are small they still result  in a minor deflection/broadening of partonic trajectories, the fragmentation appears to be unaffected.

 \begin{figure}[htb]
	\begin{minipage}{0.46\linewidth}
		\begin{center}
			\includegraphics[width=\linewidth]{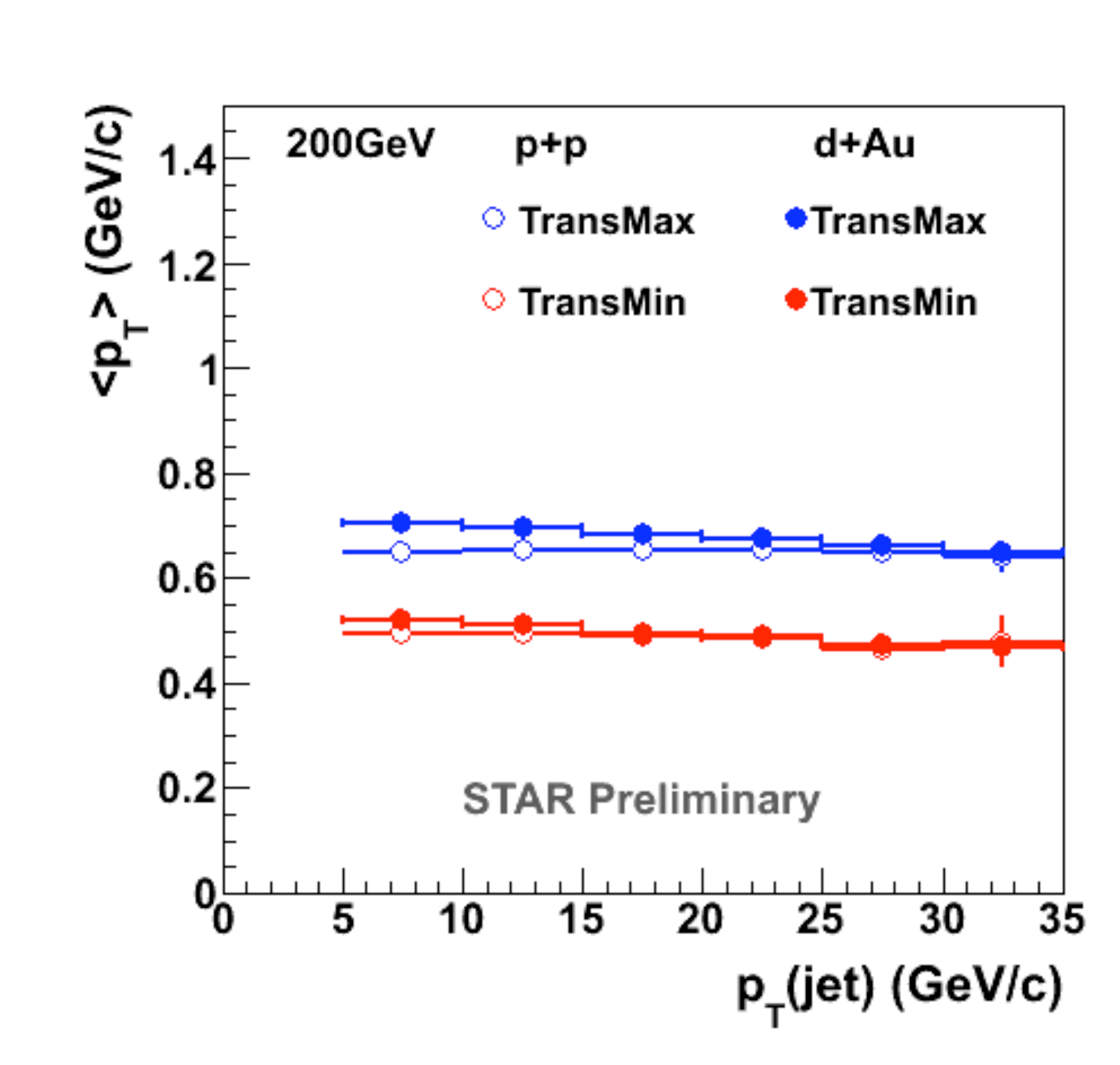}
		\end{center}
	\end{minipage}
	\begin{minipage}{0.46\linewidth}
		\begin{center}
			\includegraphics[width=\linewidth,angle=90]{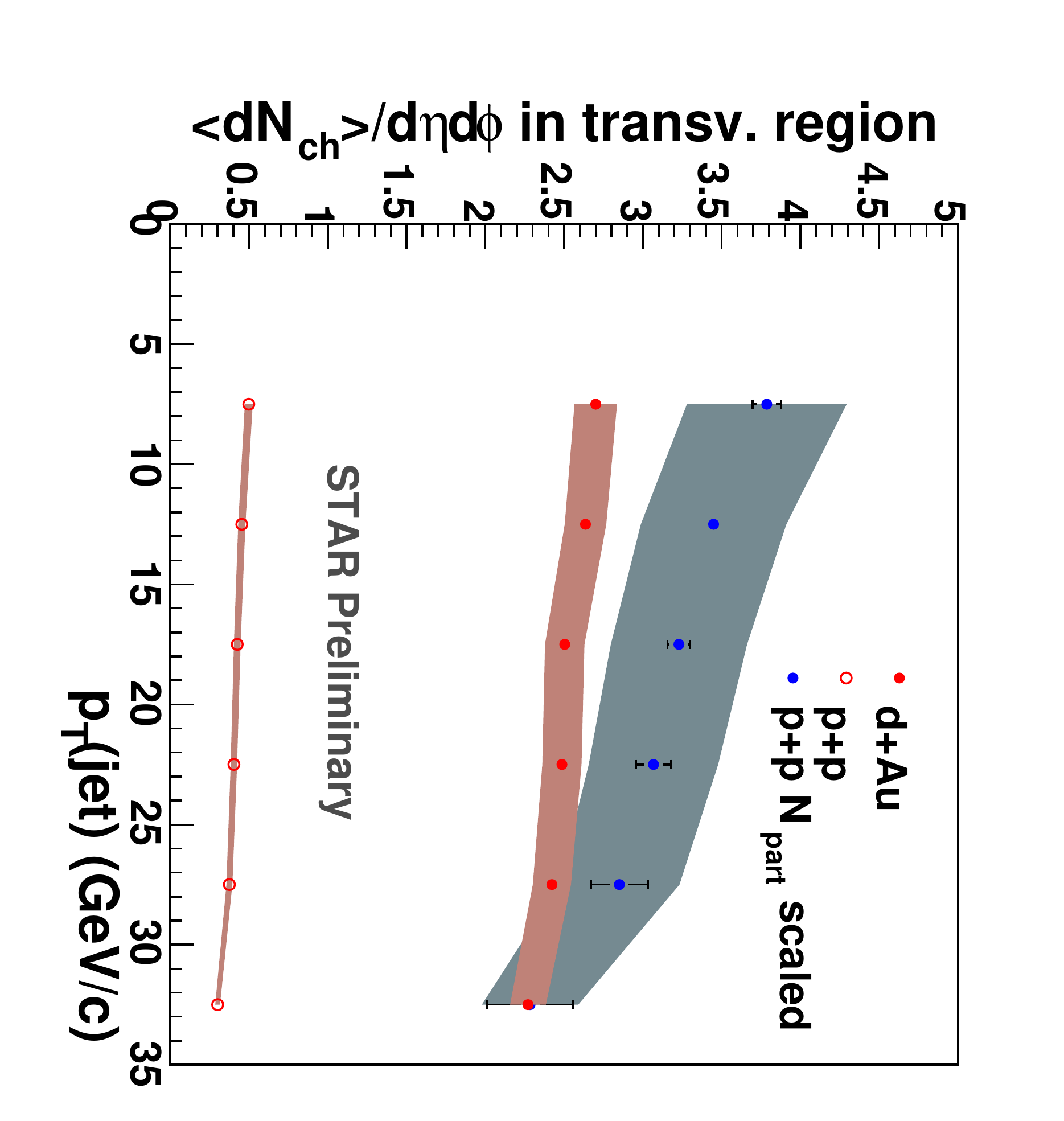}
		\end{center}
	\end{minipage}
	\caption{ Mean \pT\ (left) and mean number of charged particles per unit $\eta$ and $\phi$ (right) in the transverse regions for \pp\ and  {\it d}-Au collisions.  All data are from  $\sqrt{s_{NN}}$=200 GeV.} 
			\label{Fig:UE}
\end{figure}

The underlying event (UE) is an important element of hadronic collisions and is defined as those particles not produced in the  initial hard scatterings.   Hence it includes beam-beam remnants, particles from initial and final state re-scatterings and those resulting from soft or semi-hard multi-parton interactions, pile-up is not included in the UE definition and must be removed. In \pp\ events at RHIC the UE is small and is often neglected, however in  {\it d}-Au collisions it becomes  sizable. CDF initiated such an analysis~\cite{CDF}.   First the jets are reconstructed, next each event is split into four sections defined by their azimuthal  angle with respect to the leading jet axis ($\Delta{\phi}$). The range within $|\Delta{\phi}|$$<$60$^{0}$ is the lead jet region and an away jet area is designated for $|\Delta{\phi}|$$>$120$^{0}$. This leaves two transverse sectors of $60^{0}$$<$$|\Delta{\phi}|$$<$120$^{0}$ and $120^{0}$$<$$\Delta{\phi}$$<$-60$^{0}$. One  is  called the TransMax region and is the transverse sector containing the largest charged particle multiplicity. The second sector is termed the TransMin region. 
Two sets  analyses can then performed, a ``leading" jet study, where at least one jet is found in STAR's acceptance, and a ``back-to-back"   study which is a sub-set of the ``leading" jet collection.  This sub-set of events  has two (and only two) found jets  with $p_{T}^{awayjet}/p_{T}^{leadjet}$$>$0.7 and $|\Delta{\phi_{jet}}|$$>$150$^{0}$, this selection suppresses hard initial and final state radiation of the scattered parton. The TransMax region has an enhanced probability of containing contributions from these hard initial and final state radiation components. Thus, by comparing the TransMax and TransMin regions in the ``leading" and ``back-to-back" sets we can extract information about the various components in the UE. The properties of the UE in both \pp\ and  {\it d}-Au events  are being studied, this is the first time such an analysis has been undertaken for {\it d}-Au collisions. Since this study is preformed at mid-rapidity it is likely that there is little to no contribution from the beam-beam remnants.   Both the number of particles in and the momentum distribution of the underlying event appear to be largely independent of the leading jet's \pT\ in both \pp\ and  {\it d}-Au collisions, Fig.~\ref{Fig:UE}.  The mean transverse momentum  is  similar for \pp\ and  {\it d}-Au events in both the TransMax and TransMin regions as can be seen in Fig.~\ref{Fig:UE} left panel. Meanwhile the average number of charged particles per unit $\eta$ and $\phi$  increases by $\sim$factor 5 from \pp\ and  {\it d}-Au collisions,  right panel of Fig.~\ref{Fig:UE}. This increase in particle production is only slightly less than $N_{part}$ scaling of the \pp\ data would predict, also shown in figure~\ref{Fig:UE}.   All the results  are the same within errors for the ``leading" and ``Back-to-Back" data sets~\cite{CainesHP,Bielcikova}, which  suggests that the hard scattered partons emit  very small amounts of large angle initial/final state radiation at RHIC energies.

\section{Jets in heavy-ion collisions}

\begin{figure}[htb]
	\begin{minipage}{0.46\linewidth}
		\begin{center}
			\includegraphics[width=0.7\linewidth]{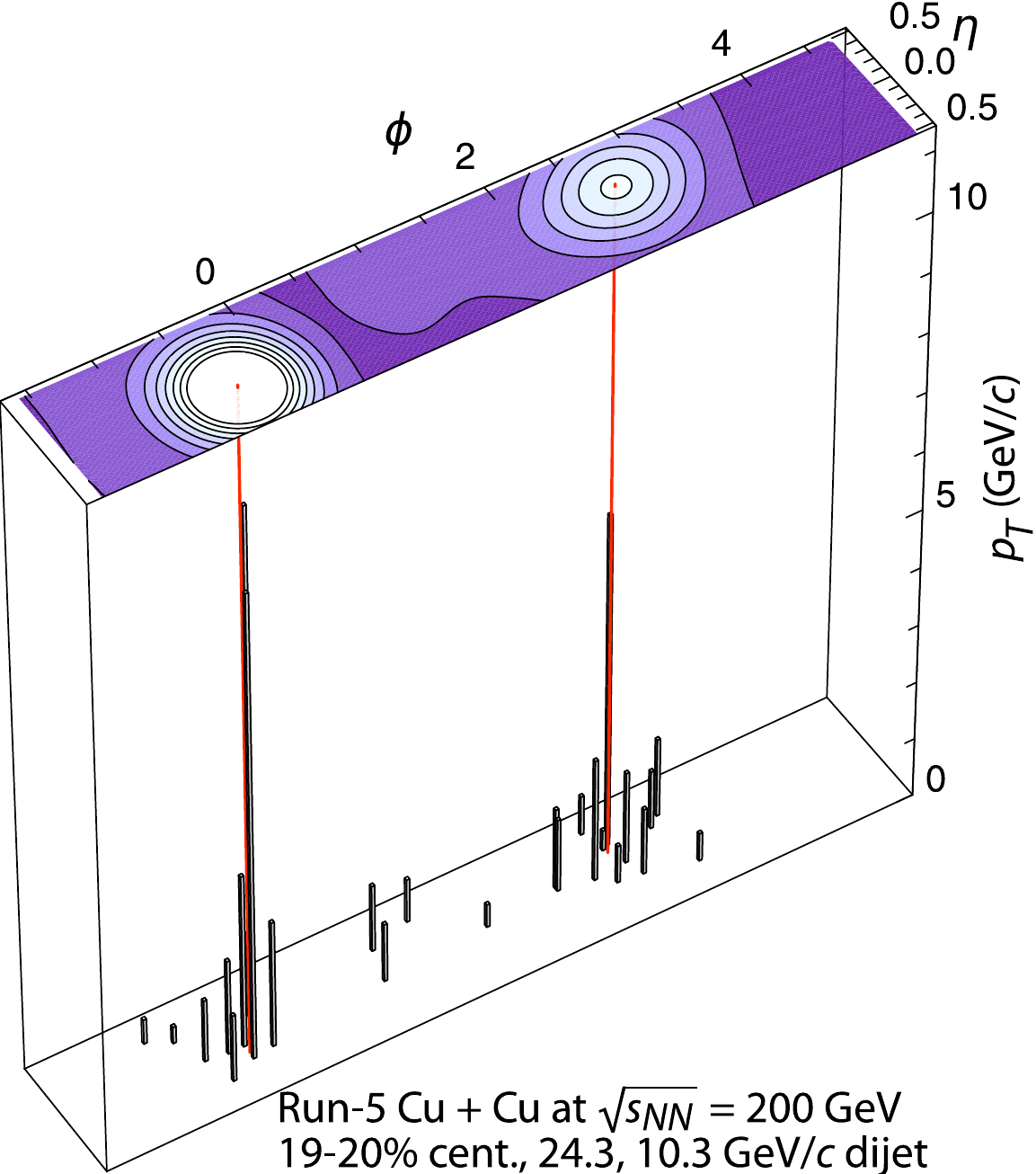}
		\end{center}
	\end{minipage}
	\begin{minipage}{0.46\linewidth}
		\begin{center}
			\includegraphics[width=\linewidth]{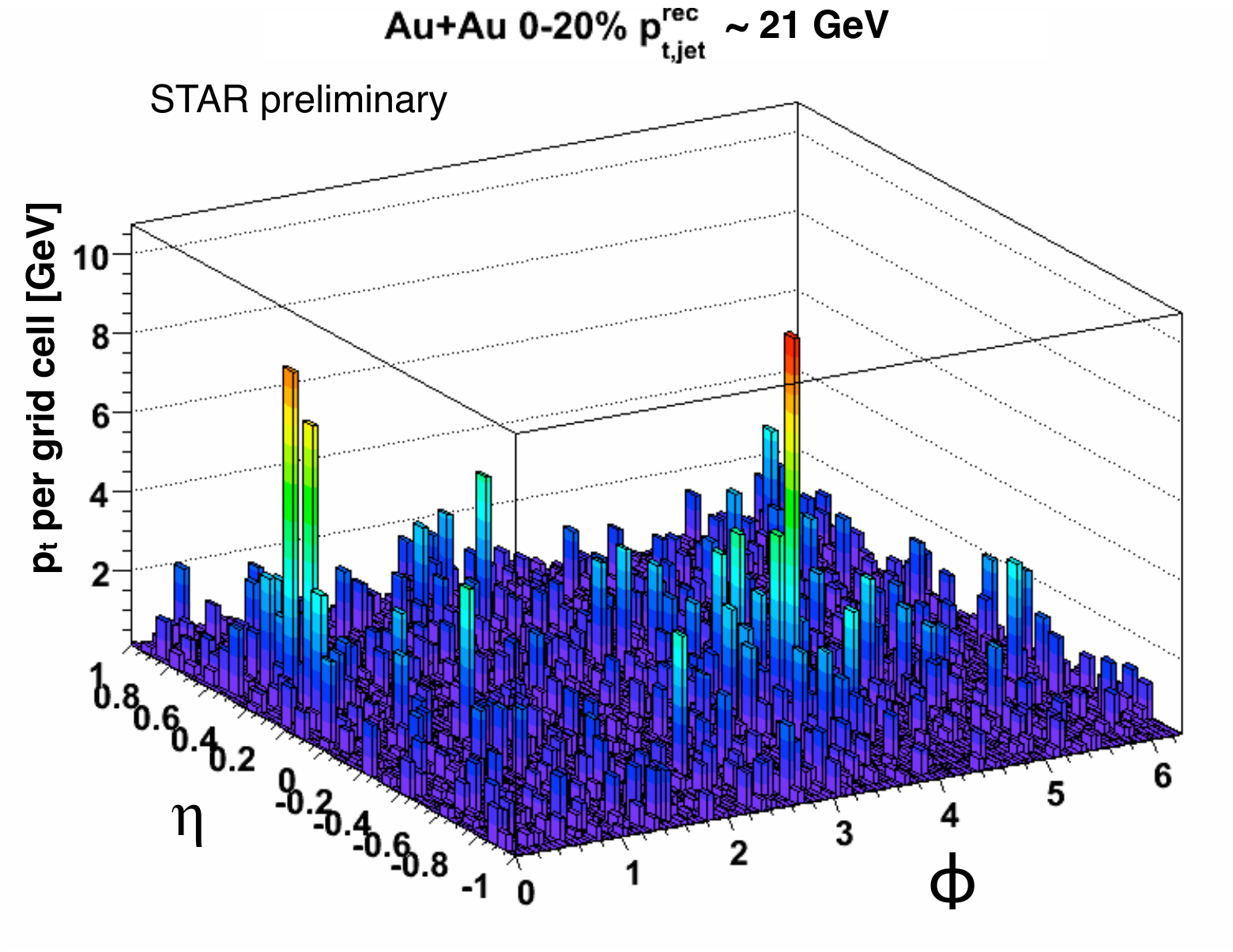}
		\end{center}
	\end{minipage}
	\caption{ Di-jets in heavy-ion events at $\sqrt{s_{NN}}$ =200 GeV Left:  PHENIX Run-5 Cu-Cu  $\sim20\%$ centrality. Charged tracks and photons are shown at the bottom by a lego plot. The distribution of the Gaussian filter output values of the event is shown at the top as a contour plot. The maxima in the filter density are reconstructed as jet axes, shown as red lines at the positions on the contour and Lego plots. Right:  STAR central Au-Au event, the summed p$_{T}$ of charged tracks and electro-magnetic calorimeter towers per grid cell are shown. Clear di-jet peaks emerge above the background.} 
			\label{Fig:JetsInAA}
\end{figure}

The presence of jets in heavy-ion events is clearly evident in Fig.~\ref{Fig:JetsInAA}, despite the significant underlying event. The left panel shows a di-jet event in the PHENIX detector from a $\sim20\%$ central Cu-Cu collision as found by the Gaussian filter algorithm~\cite{LaiDPF09}. The right panel shows a central Au-Au event in the STAR detector. Each grid cell indicates the summed \pT\ of the charged tracks reconstructed in the TPC and neutral energy recorded in the electro-magnetic calorimeter.

In order to extract information regarding jets and the interactions of hard scattered partons with the sQGP it is essential to first understand the enormous background the jet is immersed in and its fluctuations. This background is predominantly formed from the soft underlying bulk particle production and is thus strongly dependent on the centrality of the collision. Schematically, assuming all the initial partonic energy is recovered by the jet finders  the measured jet spectrum in  Au-Au collisions is:
\begin{eqnarray}
\frac{d \sigma_{AA}}{dp_{T}} = \frac{d \sigma_{AA}}{dp_{T}} \otimes F(A,p_{T})
\end{eqnarray} 

where {\it F(A, $p_{T}$)} accounts for the background and its fluctuations, which are a function of the jet area and the  \pT\ of the reconstructed jet. Initially it was assumed that these fluctuations could be accounted for by a simple Gaussian ansatz, however more detailed studies have shown this modeling to be insufficient. If the background is due to independently  emitted particles then in a fixed area the number fluctuations are  well described by a Poisson distribution, while those of the mean \pT\ result in a Gamma distribution, assuming a fixed number of particles, M~\cite{Tannenbaum}. Therefore

\begin{eqnarray}
  F(A,p_{T}) = Poisson(M(A)) \otimes  \Gamma(M(A)\langle p_{T} \rangle)
\end{eqnarray} 

\begin{figure}[htb]
	\begin{minipage}{0.46\linewidth}
		\begin{center}
			\includegraphics[width=\linewidth]{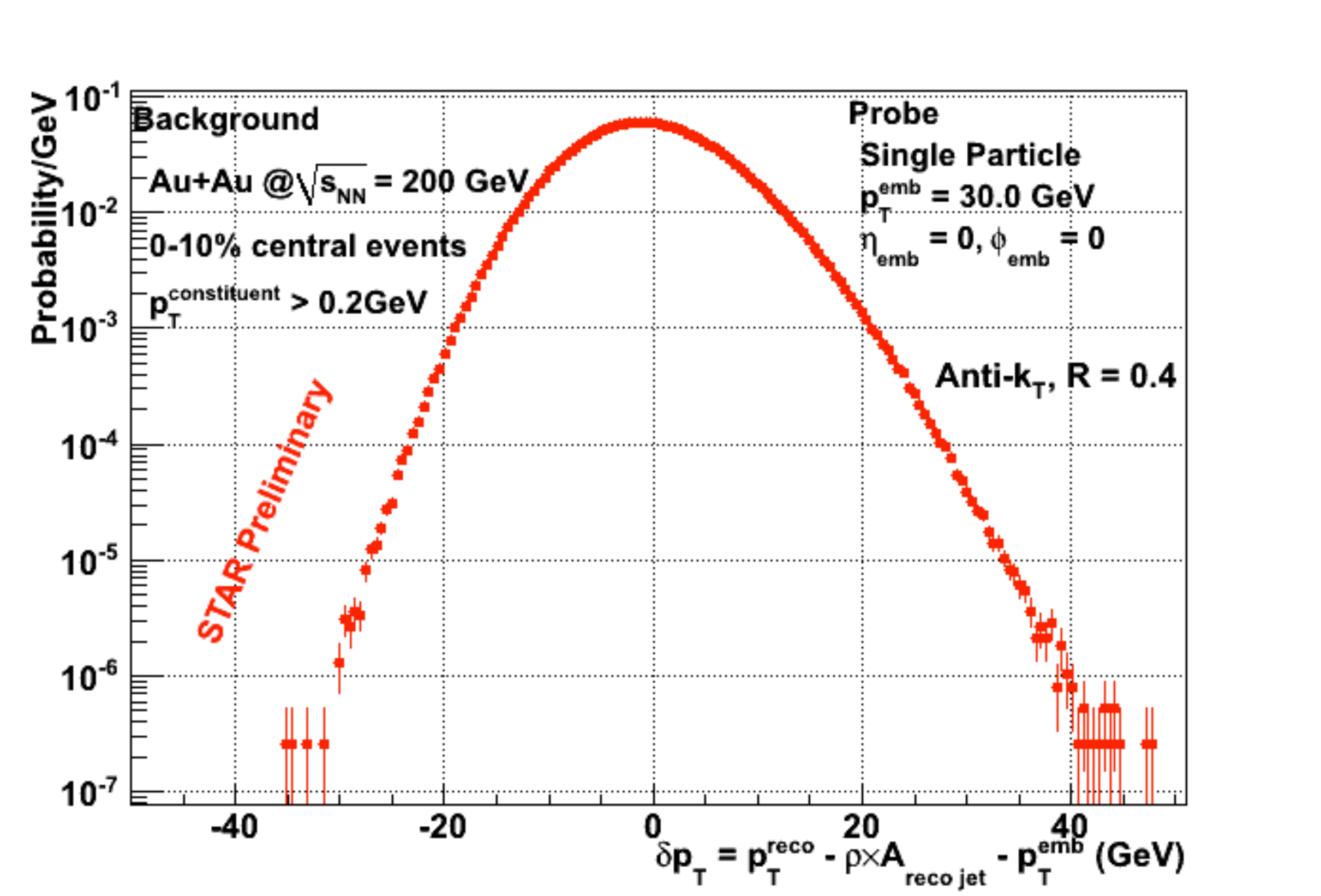}
		\end{center}
	\end{minipage}
	\begin{minipage}{0.46\linewidth}
		\begin{center}
			\includegraphics[width=\linewidth]{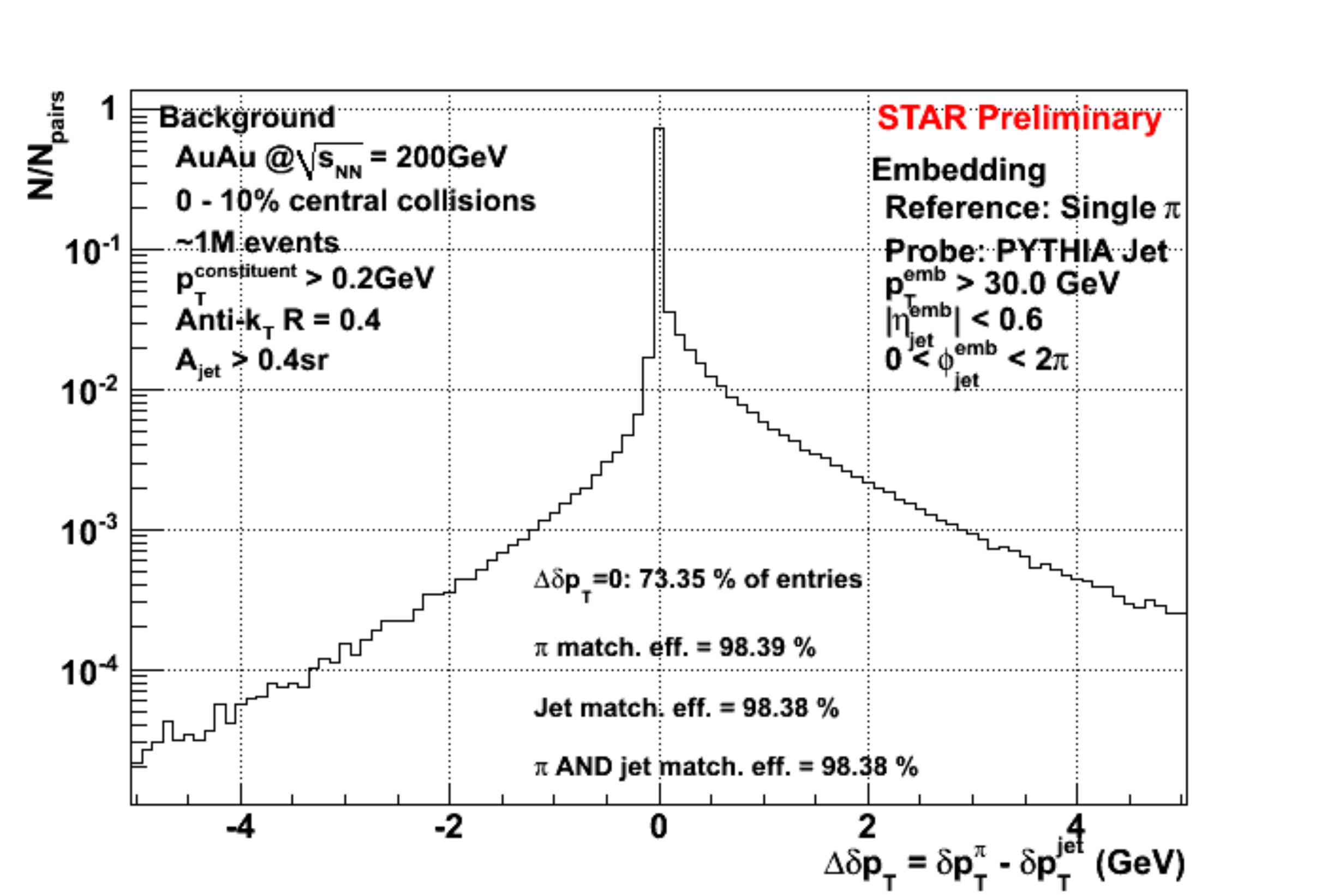}
		\end{center}
	\end{minipage}
	\caption{Left: $ \delta p_{T}$ distribution for a single pion with \pT\ = 30 GeV/{\it c}  embedded into a 0-20$\%$ central Au-Au event. Right: The event-by-event difference in $ \delta p_{T}$ for a PYTHIA embedded jet, $p_{T}>$30 GeV/{\it c}, and a single pion with same $p_T$, $\eta$ and $\phi$ as the jet.  } 
			\label{Fig:dpt}
\end{figure}

Such a  modeling of {\it F(A, $p_{T}$)} gives a good description of  the the summed \pT\ in  the random cones of area A on a toy simulation where particles with dN/d$\eta$ = 650 are thrown with a T=290 MeV. The mean of the {\it F(A, $p_{T}$)} distribution is given by $\rho A$.  $\rho$ is the median$\{p^{jet,reco}_{T,i}/A_{i}\}$ in the event and $p^{jet,reco}_{T}$ is the reconstructed jet \pT\.. When however the FastJet jet finders are used, the description is not exact, suggesting that the jet finders clustering does not occur in a truly random fashion, as should be expected.  To investigate the resilience of the jet finding in heavy-ion events to these fluctuations we are using probe embedding into real Au-Au events. A particle, or jet, of a known transverse momentum, $p_{T}^{embed}$, is embedded into an event, the Anti-K$_{T}$ algorithm run and the reconstructed jet containing the embedded probe identified. Then 
\begin{eqnarray}
  \delta p_{T} = p_{T}^\text{jet,reco} - \rho\cdot A - p_{T}^\text{embed}
\end{eqnarray}
 is calculated. The  $ \delta p_{T}$ distributions are then calculated over many events for different embedded objects. Single pions, PYTHIA jets, and qPYTHIA jets (where the fragmentation pattern is altered from that of vacuum fragmentation)~\cite{QPYTHIA} have been used as well as various p$_{T}^{embed}$. Figure~\ref{Fig:dpt} left panel shows the resulting $ \delta p_{T}$ distribution for a single 30 GeV/{\it c} pion embedded into a 0-20$\%$ central Au-Au event~\cite{BarrosPanic}.  The right panel of Fig.~\ref{Fig:dpt} shows the distribution of the  event-wise difference
\begin{eqnarray}
  \Delta\delta p_{T} = \delta p_{T}^{\pi} - \delta p_{T}^\text{jet},
\end{eqnarray}
between $\delta p_{T}$ for a  PYTHIA-generated jet with $p_T>30$ GeV/{\it c} probe and that of a pion with the same $p_T,\eta$, and $\phi$. Similar results are seen when qPYHTIA jets are used and for  lower \pT\ probes. This study reveals the Anti-K$_{T}$ response is insensitive to the fragmentation pattern of the  probe,  greater than 70$\%$ of the time  $|\Delta\delta p_{T}| < $ 200MeV{\it /c}. This is a crucial property for the jet finder used in heavy-ion jet quenching studies since we do not yet have a complete description of the fragmentation functions of partons which traverse the sQGP. Also importantly the   Anti-K$_{T}$ algorithm has been shown to respond in a predominantly geometric fashion and hence is fairly impervious to back-reaction effects~\cite{antikt}.

\begin{figure}[htb]
	\begin{minipage}{0.46\linewidth}
		\begin{center}
			\includegraphics[width=\linewidth]{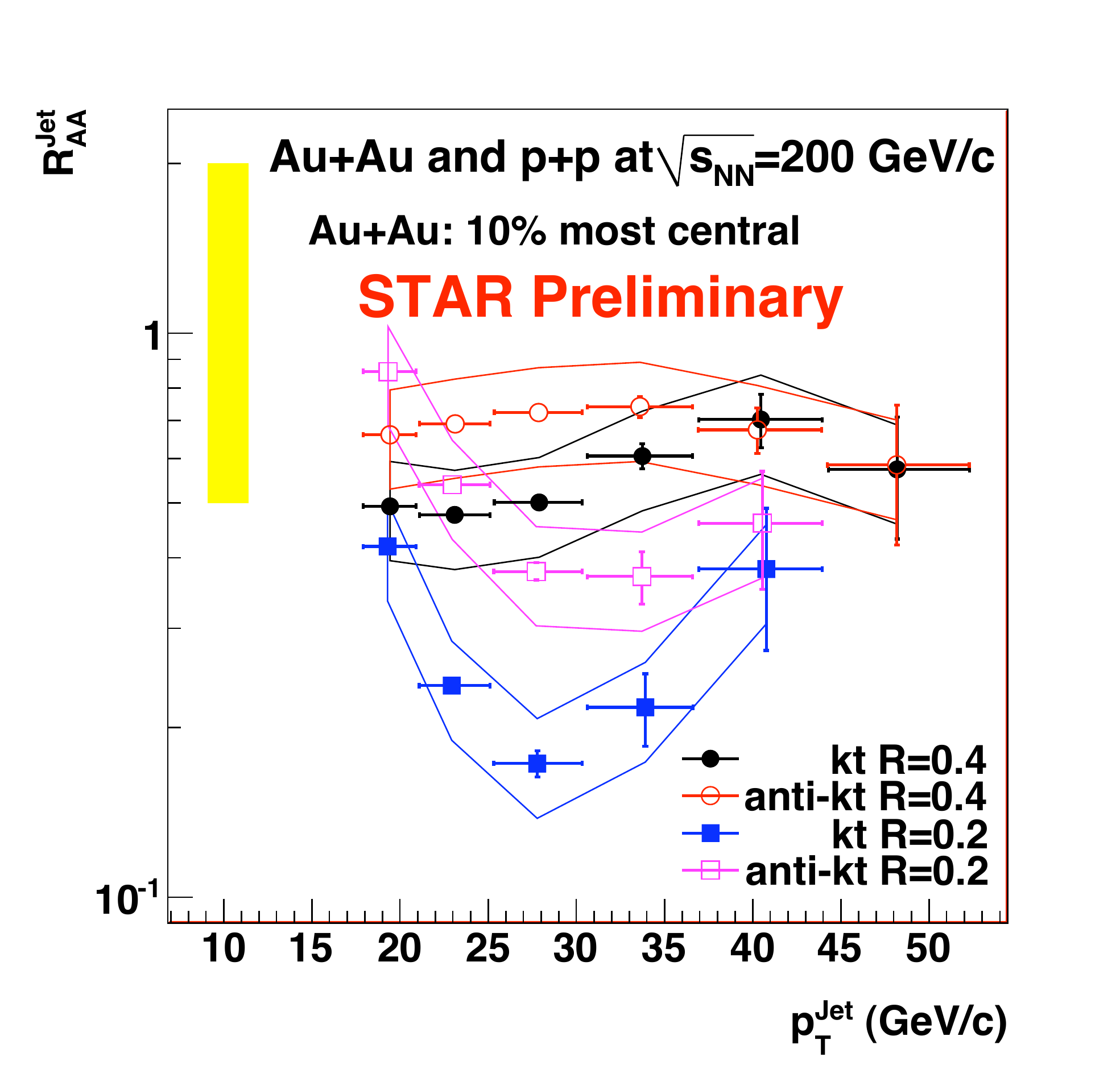}
		\end{center}
	\end{minipage}
	\begin{minipage}{0.46\linewidth}
		\begin{center}
			\includegraphics[width=0.95\linewidth]{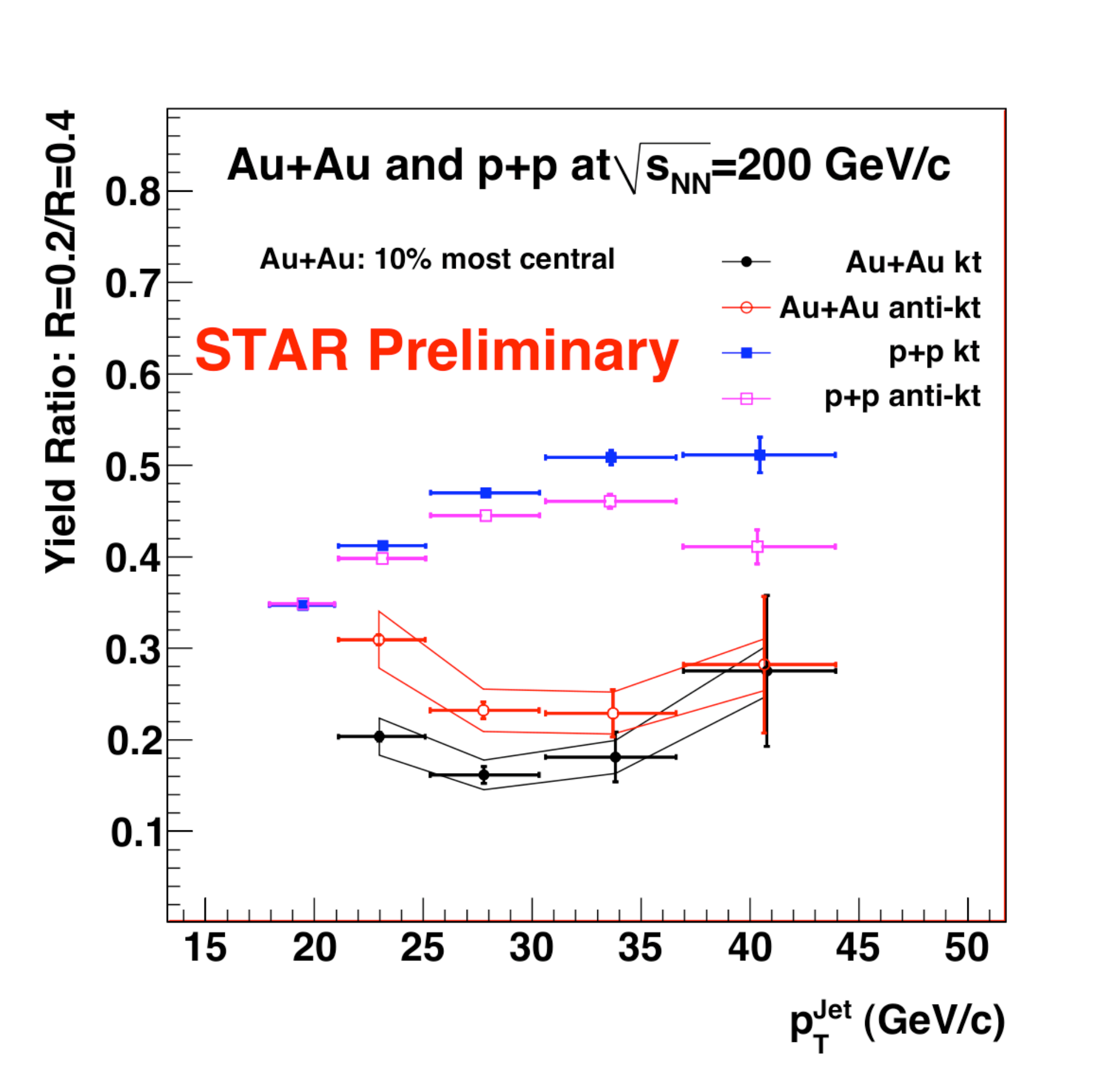}
		\end{center}
	\end{minipage}
	\caption{ Left: Jet R$_{AA}$. Right: The ratio of the number jets found with R=0.2 to those found with R=0.4 for \pp\ and Au-Au events as a function of jet $p_{T}$. } 
			\label{Fig:JetAA}
\end{figure}

For unbiased jet reconstruction one would expect the jet $R_{AA}$ to be close to unity, possible deviations might occur due to initial state effects in the Au-Au collisions. However, the left panel of Fig.~\ref{Fig:JetAA} shows that even for R=0.4 the jet R$_{AA}$ is likely below unity, due to the large systematic errors the results are just compatible with unity~\cite{QM09}. The   jet $R_{AA}$ is however significantly above that of single hadrons with \pT\ $>$20 GeV/{\it c}~($R_{AA}^\textrm{hadron}\approx 0.2$). One can also observe that there are significant differences between the results for the K$_{T}$ and Anti-K$_{T}$  algorithms which is expected given their different responses to the heavy-ion background. The ratio of the number of reconstructed jets for  R=0.2 compared to R=0.4 is less for Au-Au data than for \pp\, right panel of Fig.~\ref{Fig:JetAA}~\cite{QM09}. Taking these two results together we conclude that the jet algorithms do not recover as much of the  original partonic energy in Au-Au events as  the same algorithms and settings run on \pp\ data. Further, Fig.~\ref{Fig:JetAA}  indicates that this is likely due to the fact that particles are emitted at larger cone angles in Au-Au events compared to \pp\ events with the same jet energy, with considerable energy, even at higher jet \pT\ outside of R=0.4.

\begin{figure}[htb]
	\begin{minipage}{0.46\linewidth}
		\begin{center}
			\includegraphics[width=\linewidth]{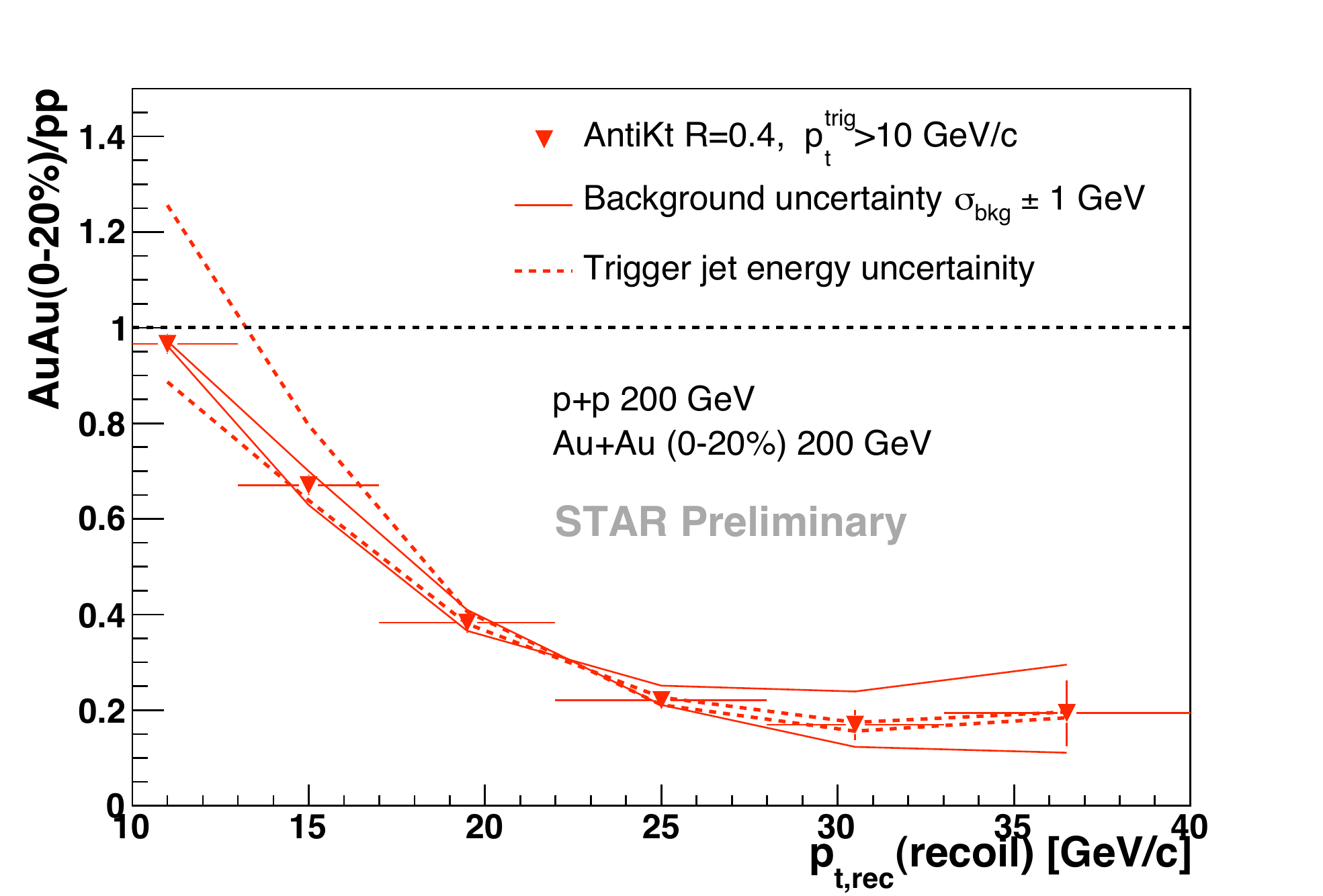}
		\end{center}
	\end{minipage}
	\begin{minipage}{0.46\linewidth}
		\begin{center}
			\includegraphics[width=0.95\linewidth]{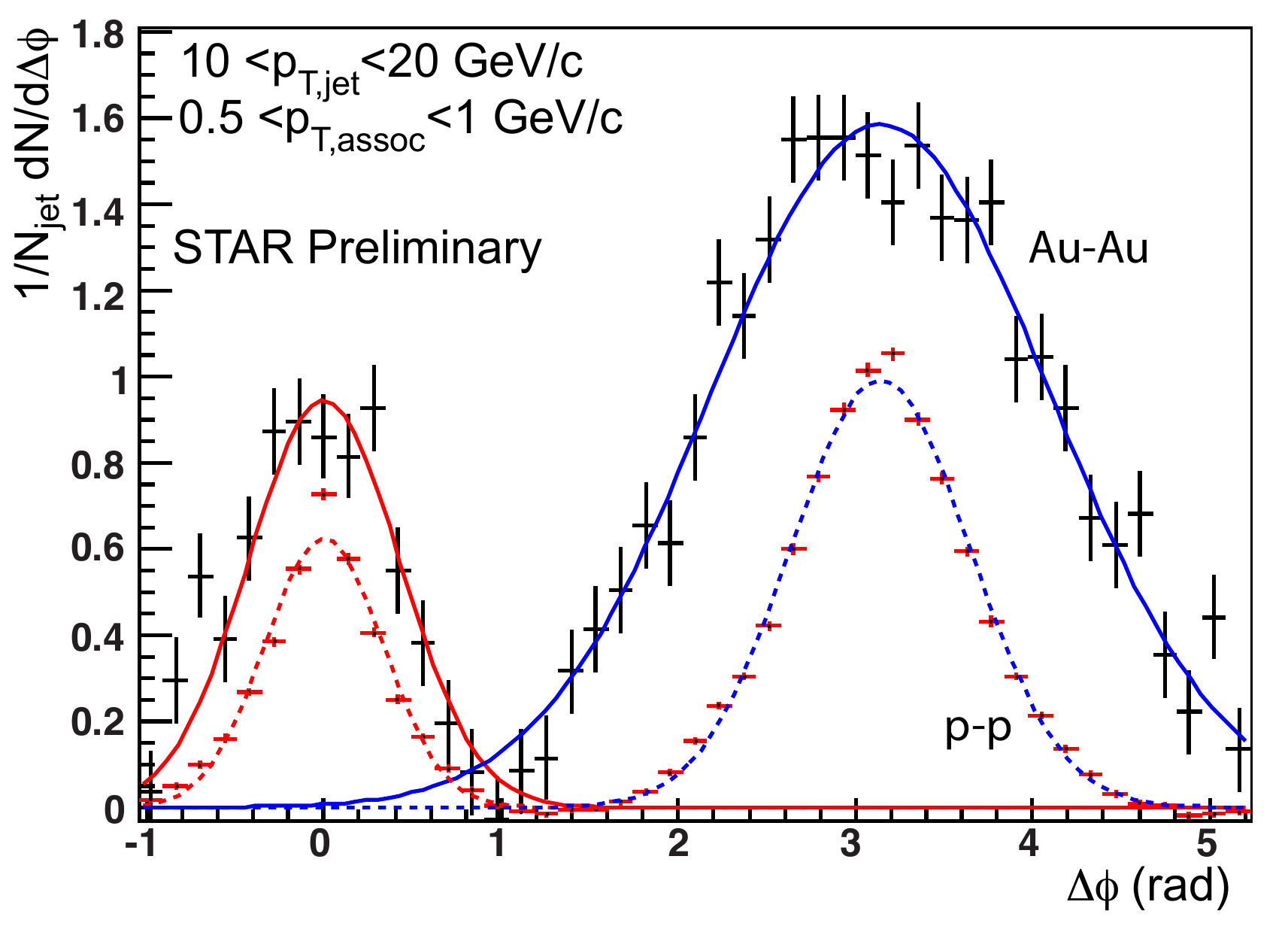}
		\end{center}
	\end{minipage}
	\caption{Left: Di-jet reconstruction probability in Au-Au over \pp\ for R=0.4 using the Anti-K$_{T}$ algorithm. The curves indicate the systematic uncertainties in the estimation of the background fluctuations (solid) and of the p$_{T}$  of the trigger jet assuming a jet resolution of 25$\%$ (dotted). Right: Jet-hadron correlations for 0.5 $ < p_{T}^{assoc}  < $ 1 GeV/{\it c} for \pp\ and Au-Au collisions.} 
			\label{Fig:diJet}
\end{figure}

Di-jet coincidence rate measurements provide us with evidence that there is a path-length dependence to the partonic energy loss. In this analysis ``trigger" jets are identified which have a reconstructed jet \pT\ $>$ 20 GeV/{\it c} when only particles with \pT\ $>$ 2 GeV/{\it c} are considered by the Anti-K$_{T}$ algorithm. These trigger jets also contain a barrel electro-magnetic calorimeter tower with $E_{tow}>$ 5.4 GeV/{\it c}. This high z fragmentation requirement biases the trigger jet to being preferentially emitted from the surface of the medium and/or to have only minimally interacted with the medium. Such a surface bias in turn maximizes the  average distance traversed by the recoil jet through the medium.  If partonic energy loss is dependent on the path-length through the medium, the recoil jet will therefore reveal a greater suppression than that  observed for the unbiased jet population. The relative probability of  reconstructing a di-jet pair in Au-Au is compared to that in \pp\ is shown in Fig.~\ref{Fig:diJet}. This relative probability is suppressed  by an approximate factor of 5~\cite{QM09di}, i.e. a much stronger rate of suppression than observed for the inclusive jets. This results supports the notion of a path length dependent energy loss term.

\begin{figure}[htb]
	\begin{minipage}{0.46\linewidth}
		\begin{center}
		\includegraphics[height=\linewidth,angle=90]{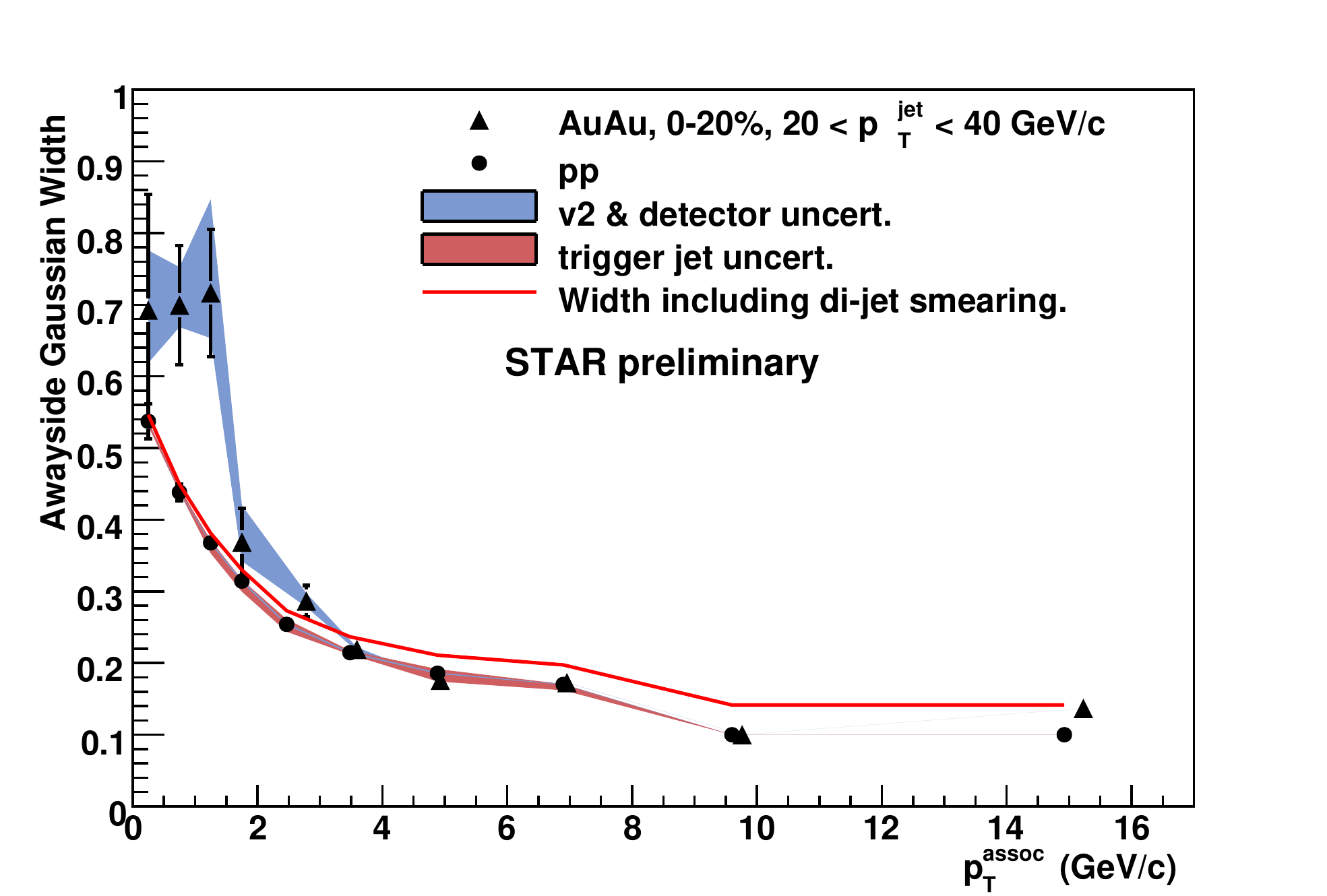}
		\end{center}
	\end{minipage}
	\begin{minipage}{0.46\linewidth}
		\begin{center}
		\includegraphics[width=\linewidth]{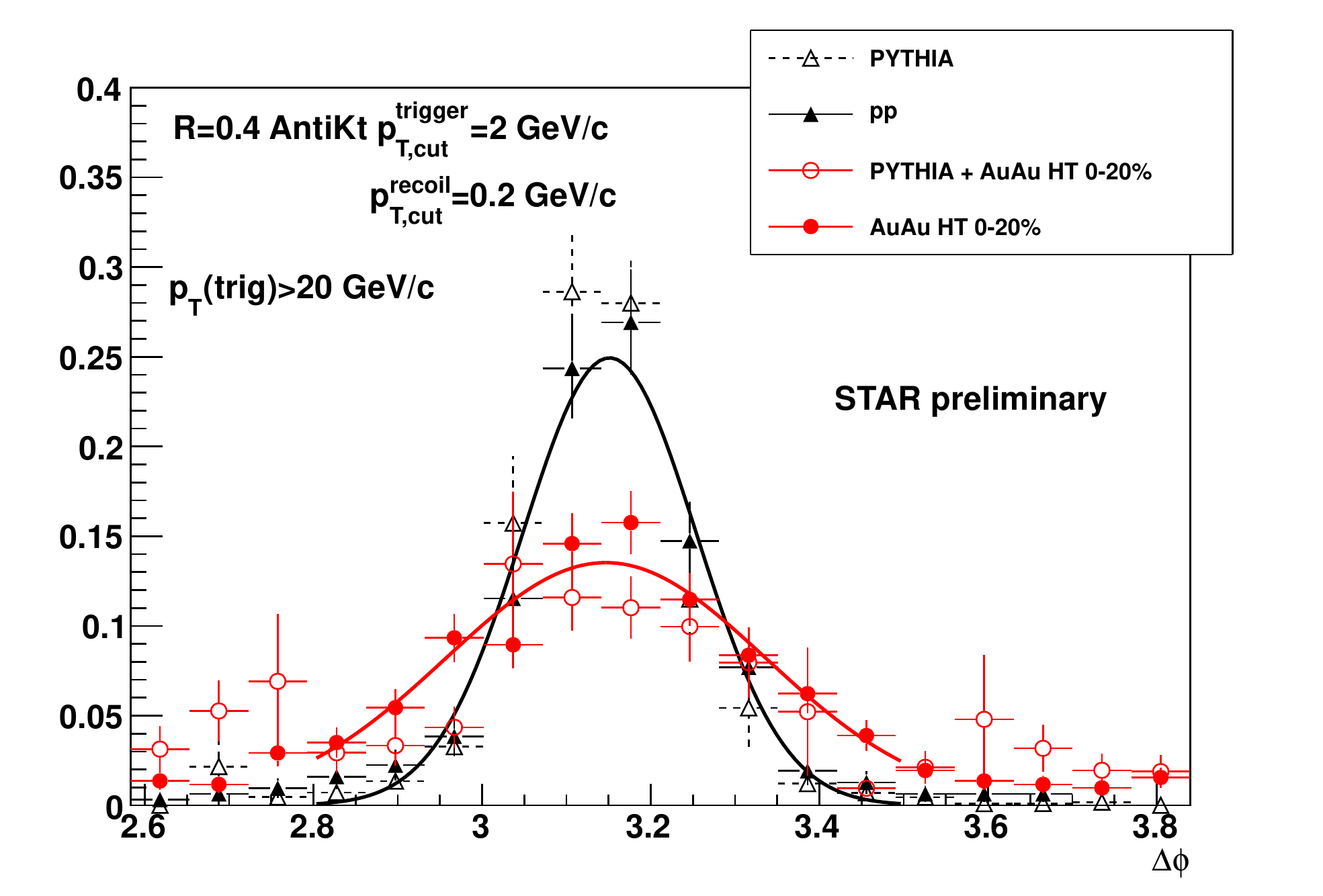}
		\end{center}
	\end{minipage}
	\caption{Left: The Gaussian widths of the away-side correlations as a function of p$_{T}^{assoc}$. Right: Di-jet $\Delta \phi$ distributions for various data sets at $\sqrt{s_{NN}}$=200 GeV collisions. Solid curves are Gaussian fits to the data.}
			\label{Fig:deltaPhi}
\end{figure}

\begin{figure}[htb]
	\begin{minipage}{0.46\linewidth}
		\begin{center}
			\includegraphics[width=\linewidth]{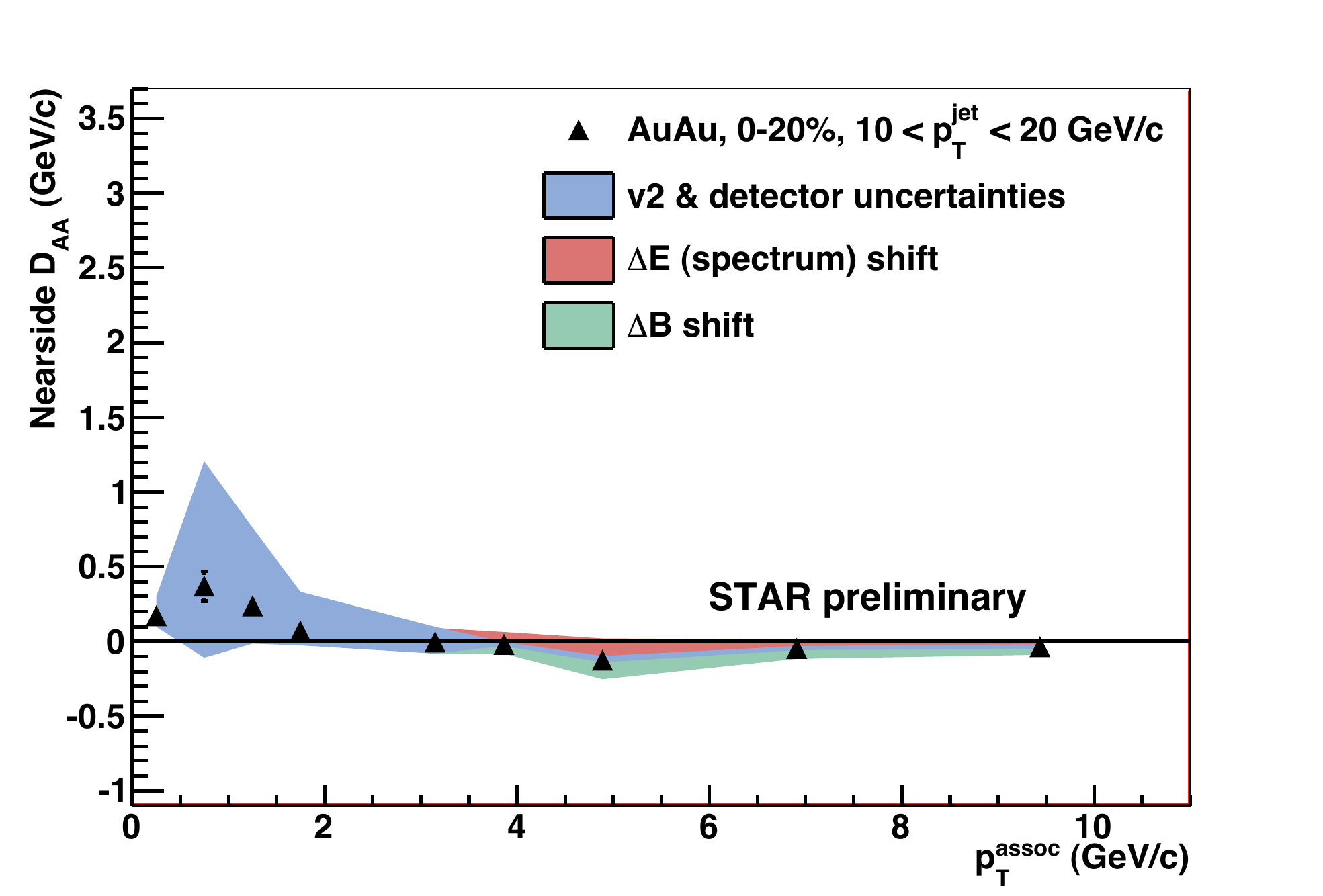}
		\end{center}
	\end{minipage}
	\begin{minipage}{0.46\linewidth}
		\begin{center}
			\includegraphics[width=\linewidth]{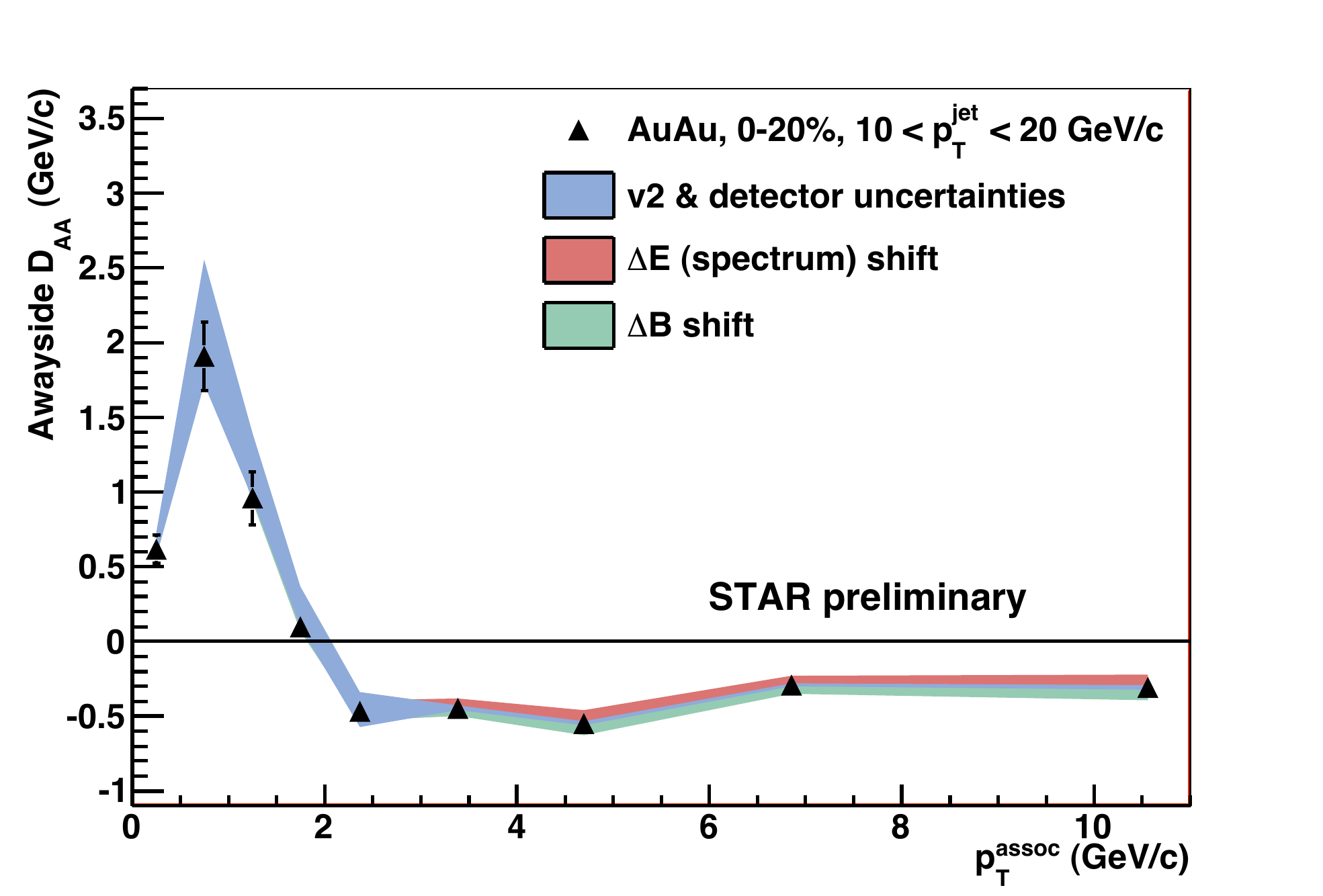}
		\end{center}
	\end{minipage}
	\caption{Jet-hadron $D_{AA}$ distributions for  the near-side (left)  and away-side (right). For details on the systematic uncertainty bands see \cite{Ohlson}.}
			\label{Fig:DAA}
\end{figure}

   To investigate further the jet broadening and softening indicated by the studies mentioned above  we turn to jet-hadron correlations.  In this analysis  a ``trigger" jet (defined as in the di-jet analysis above) is used to determine the jet axis and  the $\Delta \phi$ correlation of  {\it all} charged particles in the event relative to this axis is examined, for more details on this analysis see \cite{Ohlson}. An example of such a correlation is shown in Fig.~\ref{Fig:diJet} right panel, again a  softening and broadening of the distributions of particles from jets is indicated for low \pT\ associated.  The per trigger $\Delta \phi$ distributions for \pp\ and Au-Au event, plotted as a function of the associated charged particle \pT\ , are summarized in Figures ~\ref{Fig:deltaPhi}  and \ref{Fig:DAA}. The Gaussian widths of the away-side correlations in  \pp\ and Au-Au are shown in Fig,~\ref{Fig:deltaPhi}. The Au-Au distributions are broader than those in \pp\ for  low \pT\ associated particles, accompanied by an significant increase in the low \pT\ associated yields~\cite{Ohlson}. For high \pT\ associated particles the Au-Au recoil jet correlation width is equivalent to that of \pp\ but there is a significant reduction in the particle yield.
 Re-scattering of the initial parton could also potentially cause such a broadening rather than  a modification of the fragmentation. Therefore the di-jet  $\Delta \phi$ distributions in \pp\ and Au-Au data, PYTHIA events, and PYTHIA jets embedded into Au-Au events were studied. The results are shown in right plot of figure~\ref{Fig:deltaPhi}. The distribution is broader for the Au-Au data, however much of this broadening can be attributed to de-resolution of the jet axis due to the large underlying event, as a similar broadening is also observed in the PYTHIA+Au-Au event data. A similar result has been reported by PHENIX, who show that the $\Delta \phi$ distributions of di-jet events in Cu-Cu collisions do not vary as a function of centrality~\cite{LaiQM09}. The red curve in the left plot of Fig.~\ref{Fig:deltaPhi} indicates the expected  width of the away-side  $\Delta \phi$ distribution if the Au-Au fragmentation was \pp\-like, but with the jet axis direction smeared to reproduced the width of the $\Delta \phi$  Au-Au di-jet data. Clearly such a smearing cannot fully explain the observed broadening, and it also does not explain the enhanced low \pT\ yields.
 
The  integrated yield difference, ($D_{AA} = Yield_{AA}\times\langle p_{T}^{assoc}\rangle-Yield_{pp}\times\langle p_{T}^{assoc}\rangle$) of the near- and away-side correlations as a function of p$_{T}^{assoc}$ are plotted in Fig.~\ref{Fig:DAA}. As expected, the ``surface" bias of the trigger causes the near-side $D_{AA}$ to be consistent with zero for all  p$_{T}^{assoc}$. This means that there is an approximate energy balance, and a similarity of the associated \pT\  particle distributions for Au-Au and \pp\ data for the trigger jet. The away-side data, Fig.~\ref{Fig:DAA} right panel, reveals  that the low \pT\ hadron enhancement in the Au-Au data is approximately matched by the high \pT\ associated particle suppression. This  suggests that the broadening and softening observed in the away-side correlation data is indeed due to a modification of the partonic fragmentation and not from residual soft  background particles.

\begin{figure}[htb]
	\begin{minipage}{0.46\linewidth}
		\begin{center}
			\includegraphics[width=\linewidth]{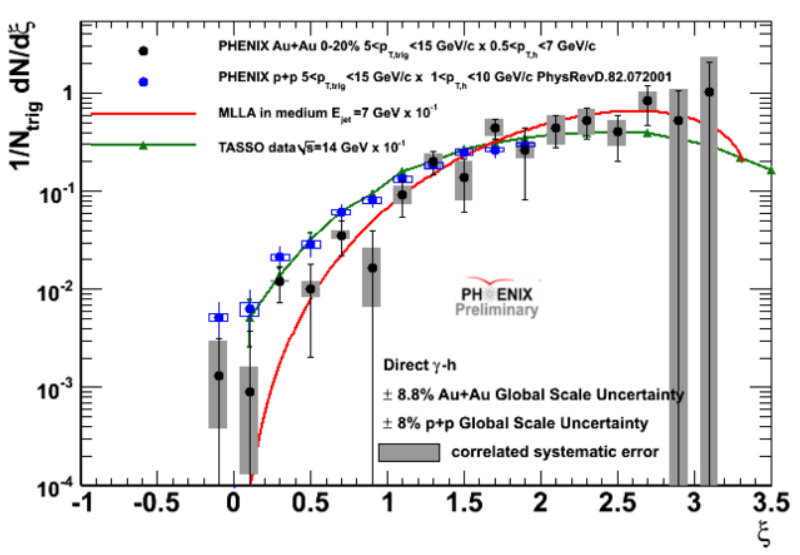}
		\end{center}
	\end{minipage}
	\caption{$\xi$ distributions from PHENIX $\gamma$-hadron correlations in Au-Au data (black circles) and \pp\ data (open blue circles) compared to TASSO data (green triangles) and MLLA in medium prediction (red curve).}
			\label{Fig:gammaJet}
\end{figure}

To remove the surface bias of the trigger object  introduced in the di-jet and jet-hadron analyses discussed above PHENIX have been investigating $\gamma$-hadron correlations.  Direct photon-hadron correlations are an ideal channel for studying energy loss since direct photons do not interact via the strong force and hence traverse the sQGP unmodified. At leading order pQCD, direct photons are produced from a Compton scattering of $q + g \rightarrow q + \gamma$ or quark annihilation $q + \bar{q} \rightarrow g + \gamma$. To conserve energy and momentum a matching recoil jet is also produced. The energy of the photon can then be used as a proxy for the jets initial energy and the fragmentation function of the recoil jet can be calculated.  $\gamma$-hadron $\Delta \phi$ correlations have been measured in both \pp\ and Au-Au collisions~\cite{ConnorsHP}, an isolation cut is used around the trigger photon to reduce contamination from $\pi^{0}$ and fragmentation photons. A fragmentation function is then deduced for the recoil jet correlation at  $| \Delta \phi - \pi | \le \pi/2$. The resulting distributions as a function of $\xi  = -ln(x_E)$ where $x_E = p_{T}^{hadron}cos(\Delta \phi)/p_{T}^{photon}$
 are shown in Fig.~\ref{Fig:gammaJet} for \pp\ and central Au-Au collisions. The preliminary Au-Au and  published \pp\ data are plotted and compared to the TASSO measurement~\cite{Tasso}  and a Modified Leading Logarithmic Approximation (MLLA) in medium prediction~\cite{MLLA}.  The TASSO data and MLLA curve have been arbitrarily scaled down by a factor of ten to account for the limited PHENIX $\eta$ acceptance as in~\cite{Adare}  The shape of the isolated \pp\ data are in good agreement with the TASSO measurement of the quark fragmentation function from $e^{+}+e^{-}$ collisions. The
Au-Au results show depletion at low $\xi$ and possible enhancement at high $\xi$ compared to the \pp\ data. These results again indicate that the energy lost by high \pT\ partons reappears as soft hadrons correlated with the initial parton's path.

\section{Summary}

In conclusion, both PHENIX and STAR are making quantitative steps in our understanding of jet production and fragmentation in \pp\, {\it d}-Au, and Au-Au collisions at RHIC at $\sqrt{s_{NN}}$ = 200 GeV over a wide kinematic range.

Both the jet and di-jet cross-section in \pp\ collisions are well described by next-to-leading order calculations once hadronization and the effects of the underlying event are taken into account. PYTHIA simulations reproduce the measured  distributions of the fragmentation products  even at large jet resolution parameters indications that NLO corrections, beyond those implemented in PYHTIA, are small.

Jet production in {\it d}-Au collisions is slightly suppressed, particularly in central {\it d}-Au events when compared to binary scaled \pp\ or peripheral  {\it d}-Au data. Together with the k$_{T}$ and j$_{T}$ measurements as a function of jet \pT\ this indicates that there are small cold nuclear matter effects present but that these do not affect the shape and distribution of the fragmentation particles produced.

Underlying event measurements of the  \pp\ and  {\it d}-Au data show no significant changes as a function of jet \pT\. The mean number of charged particles produced in the transverse region approximately scales with the number of participants in the event, while the mean \pT\ of these particles remains constant
between \pp\ and  {\it d}-Au data.

Our understanding of the background, and most importantly its fluctuations,  in heavy-ion events has significantly improved. The Gaussian ansatz of the fluctuations has been shown to be incorrect, they are more closely reproduced from a folding of a Gamma function with a Poisson that depend on the  jet area, multiplicity of the background and its mean \pT. It has also been shown that the Anti-K$_{T}$ algorithm's response to the background and its fluctuations is largely independent of the fragmentation pattern of the jet.

Using a jet resolution parameter of R=0.4 the measured jet cross-section in central Au-Au collisions does not binary scale compared to \pp\ data, the jet R$_{AA} <1 $. This reveals that the lost partonic energy is spread to radii beyond R=0.4. Further the Au-Au R=0.2/R=0.4 ratio as a function of jet \pT\ is lower than that measured in \pp\ showing this broadening is  there for all radii. The di-jet reconstruction probability in Au-Au collisions is suppressed as would be expected if the partonic energy loss is pathlength dependent. 

Both jet-hadron and  direct photon-hadron correlations  indicate an enhanced production of hadrons a low \pT\ compared to \pp\ baseline measurements which appear to compensate the suppressed particle production at high \pT. However, the high \pT\ associated particles while suppressed  in number compared to \pp\ data do not reveal any broadening in the jet-hadron correlations. This is in agreement with a scenario where the scatted parton loses energy in the sQGP but then fragments outside of the medium as it would in vacuum albeit with a reduced energy. The di-jet $\Delta \phi$ distributions indicate no obvious deflection of the parton's path although significant path length dependent energy is lost as it traverses the medium.

All of the measurements made in heavy-ion collisions show that energy lost by high $Q^{2}$ scattered partons re-appears as soft particle production, with properties similar to that of the bulk, that remains largely correlated to the jet axis.

\bigskip 

\end{document}